\documentclass[aps,prl,twocolumn,amsmath,amssymb,nofootinbib,superscriptaddress,floatfix]{revtex4-1}

\usepackage{amsmath}
\usepackage{amssymb}
\usepackage{amsthm}

\usepackage[pdftex]{color} 
\usepackage{graphicx}% Include figure files
\usepackage{dcolumn} % Align table columns on decimal point
\usepackage{bm} % bold math
\usepackage{float} % allow figure being shown in same page
\usepackage[driverfallback=dvipdfm]{hyperref} % So that it can be run with pdftex
\usepackage{longtable}
\usepackage{ulem}   % to strike things out
\newcommand{\bs}[1]{{\boldsymbol{#1}}}

\begin{document}

\author{Jian Wang}
\affiliation
{
Department of Physics,
Emory University, Atlanta, Georgia 30322, USA
}

\author{Luiz H. Santos}
\affiliation
{
Department of Physics,
Emory University, Atlanta, Georgia 30322, USA
}

\title{ 
Classification 
of Topological Phase Transitions 
and van Hove Singularity Steering Mechanism in Graphene Superlattices
}

\begin{abstract}
We study quantum phase transitions 
in graphene superlattices in external magnetic fields,
where a framework is presented
to classify multiflavor Dirac fermion critical points
describing hopping-tuned topological phase transitions of integer and fractional Hofstadter-Chern insulators.
We argue and provide numerical support for
the existence of transitions that can be explained by a nontrivial interplay of Chern bands and van Hove singularities near charge neutrality.
This work provides a route to critical phenomena beyond conventional quantum Hall plateau transitions.
\end{abstract}

\date{\today}

\maketitle

Chern bands\cite{TKNN1982,Haldane1988}
are the building blocks of the Hofstadter spectrum\cite{Hofstadter1976} when a large magnetic flux (of order $\phi_{0} = h/e$) penetrates the unit cell of the 2D lattice.
They give rise to quantum Hall phases beyond the Landau level (LL) paradigm,
which has attracted considerable interest\cite{Neupert-2011,Sheng-2011,Tang-2011,Sun-2011,Regnault2011}.
Rapid progress in the fabrication of superlattices with nanometer scale unit cells has led to the experimental realization
of
integer\cite{Dean2013,Ponomarenko2013,Hunt2013,forsythe-NatureNanoTech2018} and 
fractional\cite{Spanton2018} Hofstadter-Chern insulators (IHCI and FHCI),
thereby opening remarkable prospects to explore the nontrivial interplay of lattice effects and electronic topology that is inaccessible in regular 2D lattices.

Topological ground states supported in Chern bands have been broadly studied using different approaches including numerical methods\cite{Neupert-2011,Sheng-2011,Tang-2011,Sun-2011,Regnault2011,Liu-PRL2012,Wu-PRB2012,Lauchli-PRL2013}, composite fermions\cite{Jain1989,Lopez1991,Kol1993,Moller2015,murthyshankar2012,Sohal-2018} and
Lieb-Schultz-Mattis type constraints.\cite{Lu-Ran-Oshikawa2020}
On the other hand, the fundamental influence of lattice parameters on topological phase transitions (TPTs) in IHCIs and FHCIs has received 
significantly less attention.\cite{pfannkuche1997,sato-PRB-2008,Lee-PRX-2018}
The complexity of the Hofstadter spectrum and the finite bandwidth of Chern bands that reflects their dependence on the lattice parameters and on the intracell magnetic flux appears to stand in the way of an overarching understanding of lattice-tuned TPTs, which are distinct from plateau transitions tuned by the magnetic field.\cite{Jain-Kivelson-Trivedi-1990,Kivelson-Lee-Zhang-1992}

In this Letter, we provide a 
classification of TPTs in IHCIs and FHCIs and present a mechanism for quantum criticality tuned by lattice parameters with a fixed background magnetic field. 
Numerical studies\cite{pfannkuche1997,Lee-PRX-2018} strongly support the existence of continuous TPTs tuned by the amplitude of a square lattice
weak potential projected on the lowest LL.
This work, on the other hand, employs an effective tight-binding description (i.e. ``strong" potential) of a honeycomb superlattice with the magnetic field incorporated via Peierls substitution and discusses topological transitions tuned by hopping amplitudes of the lattice. 
Graphene superlattices realized via nanolithography
~\cite{Gibertini2009,Singha2011,Soibel_1996,N_dvorn_k_2012,Wang2016,Wang2018}
not only provide a motivation for this study but also offer promising test beds of these ideas.

The main results presented in this Letter are as follows: 
(1)
We show that hopping-tuned TPTs on the honeycomb lattice with a fixed rational intracell magnetic flux $\phi = (p/q)\phi_{0}$ are characterized by $q$ Dirac fermions (DFs) located in high-symmetry momenta of the magnetic Brillouin zone.
The number of DF flavors and their momentum space distribution are derived analytically from a nontrivial function that implicitly sets the momentum dependence of \textit{all} the Chern bands of the spectrum. 
(2) We establish a surprising connection
between van Hove singularities(VHSs)\cite{vHS-1953} and the 
onset of TPTs
near charge neutrality.
(3) This nonperturbative analysis is extended to hopping-tuned FHCI transitions described by composite fermions\cite{Jain1989,Lopez1991,Kol1993,Moller2015,murthyshankar2012,Sohal-2018} in partially filled Chern bands.

\begin{figure}[htbp]
\centering
\includegraphics[width = .48\textwidth]{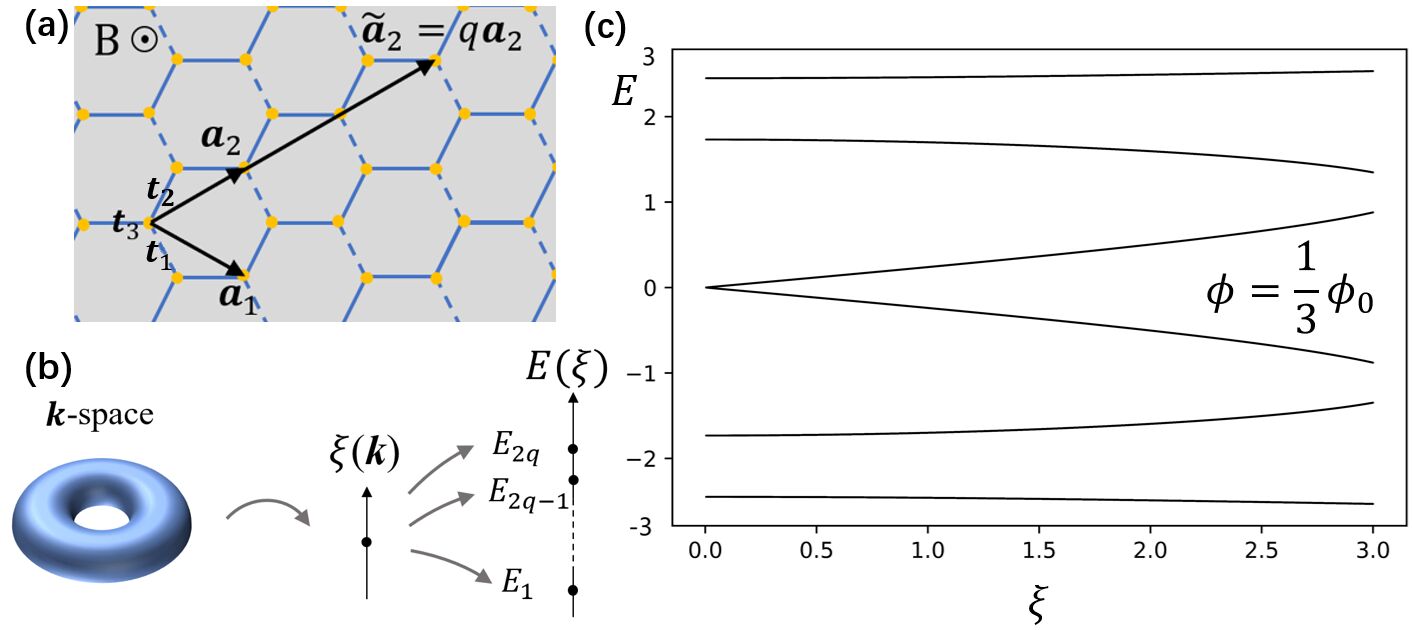}
\caption{Parameterization of the Hofstadter-Chern bands by the Thouless function. (a) Honeycomb superlattice with lattice constant $a$ in the nanometers and magnetic unit cell $q$ times extended along $\bs{a}_{2}$.
(b) Momentum dependence on the Thouless function $\xi$. (c) Spectrum as function of $\xi$ for $\phi = (1/3)\phi_{0}$.}
\label{fig: lattice}
\end{figure}

Our setting is a honeycomb superlattice in a external perpendicular magnetic field, $B = \partial_{x}A_{y} - \partial_{y}A_{x}$, described by the single-particle nearest neighbor effective Hamiltonian
\begin{eqnarray}
\label{eq: TB Hamiltonian 1}
H = -\sum_{<\bs{r},\bs{r}'>} 
t_{\bs{r},\bs{r}'}
\,
\,
e^
{\mathrm{i}\,
\frac{2\pi}{\phi_{0}}
\,
\int_{\bs{r}}^{\bs{r'}}\,d\bs{x}\cdot\bs{A}(\bs{x})
}
\,
a^{\dagger}_{\bs{r}}
b_{\bs{r}'}
+
\mathrm{H.c.}
\end{eqnarray}
$a^{\dagger}_{\bs{r}} = a^{\dagger}_{m,n}$ and $b^{\dagger}_{\bs{r}} = b^{\dagger}_{m,n}$ are spin polarized fermionic creation operators on the two sublattices, $\bs{r} = m\bm{a}_1 +  n\bm{a}_2 $, $m, n \in \mathbb{Z}$ is the lattice vector with basis vectors 
$\bs{a}_1 = a\,(3/2,-\sqrt{3}/2)$ 
and 
$\bs{a}_2 = a\,(3/2,\sqrt{3}/2)$, 
and
$t_{\bs{r},\bs{r}'} = \{ t_1, t_2, t_3 \}$
are nearest neighbor real hopping elements, as shown in (a) in Fig.\ref{fig: lattice}.

Working in the gauge
$\bs{A}=\hat{y}(x+\sqrt{3}y)B$
with rational flux $\phi = B\,{\frac{\sqrt{3}}{2}}a^{2} = (p/q)\phi_0$ ($p, q \in \mathbb{Z}_{+}$ and coprime), we introduce the magnetic unit cell containing $2q$ sites as in Fig.\ref{fig: lattice}(a), which leads to the $\bs{k}$-space Hamiltonian \cite{supplmat}
$
H = 
-
\sum_{\bs{k} \in \textrm{MBZ}}
\psi^{\dagger}_{\bs{k}}
{\tau_{1}\otimes h_{\bs{k}}}
\psi_{\bs{k}}
$,
where
$
\bs{k}
=
(k_1,k_2)
\equiv
k_1\,
\tilde{\bs{g}}_{1}
+
k_2\,
\tilde{\bs{g}}_{2}
$
is the momentum expanded along the reciprocal lattice vectors $\tilde{\bs{g}}_{1,2}$.

Aiming at a nonperturvative description of the Chern bands beyond the isotropic lattice $t_1=t_2=t_3$\cite{rammal1985,bernevig-IJMPB-2006,Agazzi2014}, we establish the spectral function 
$
\mathcal{P}(E) 
=
\textrm{det}
\left(
E\,\mathrm{I} - H 
\right)
$,
\begin{subequations}
\label{eq: implicit momentum dependence}
\begin{equation}
\label{eq: E polynomial}
\begin{split}
\mathcal{P}(E) 
&\,= 
\sum^{q}_{n=1}\,
a_{n}(\{ t_i \})\,E^{2n} 
- 
\xi^{2}(\{ t_i \}, k_1, k_2)
\,,
\end{split}
\end{equation}
\begin{equation}
\label{eq:  xi def}
\xi(\{ t_{i} \}, k_1, k_2)
=
|
t^{q}_{1}
\,
e^{\mathrm{i} q\,k_1 -\mathrm{i}\pi\,(q-1)}
+
t^{q}_{2}
\,
e^{\mathrm{i} k_2}
+
t^{q}_{3}
|
\geq 0
\,.
\end{equation}
\end{subequations}
Equation \eqref{eq: implicit momentum dependence}
encodes a remarkable property of the Hofstadter spectrum [originally noticed by Thouless in a different context\cite{Thouless-1983} (see also\cite{rammal1985})],
namely, that the momentum dependence of the bands is ``compressed" in a single function $\xi(\bs{k})$, i.e.
$E_{\alpha}(\bs{k}) = E_{\alpha}[\xi(\bs{k})]$ for $\alpha = 1, ..., 2q$.
Figure. \ref{fig: lattice}(b), (c) shows how the energy bands depend on the ``Thouless function" $\xi$, which we notice is related to the graphene band \cite{graphene-RMP} upon the replacements
$(k_1,k_2) \rightarrow [q\,k_1 {-\pi(q-1)}, k_2]$
and $\{t_{i}\} \rightarrow \{t^{q}_{i}\}$.

\begin{figure}[htbp]
\centering
\includegraphics[width = .48\textwidth]{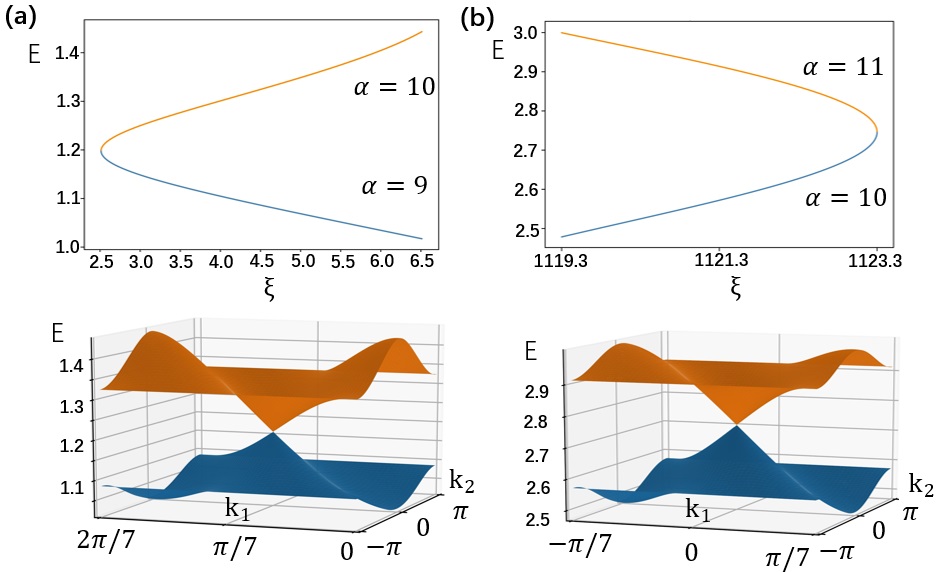}
\caption{
TPTs of the $\phi = (1/7)\phi_0$ lattice ($\alpha$ denotes band index).
(a) At $(t_1,t_2,t_3) = (1.24,1,1)$, $7$ Dirac cones (only one shown) form at $\bs{k}^{(n)}_{min} = (-\pi/7 + 2\pi n /7,0)$, $n = 0, ..., 6$.
(b) At $(t_1,t_2,t_3) = (2.73,1,1)$, $7$ Dirac cones (only one shown) form at $\bs{k}^{(n)}_{max} = (2\pi n /7,0)$, $n = 0, ..., 6$.
}
\label{fig: band touching}
\end{figure}

\noindent
\textbf{IHCI transitions - } 
We now establish a classification of TPTs in the parameter space $(t_1,t_2,t_3).$
On general grounds, consider
a TPT tuned by the hopping parameters where two Chern bands touch at
$(\xi_{F}, E_{F})$,
where 
$\xi_{F}\neq 0$ and
$E_{F} \neq 0$ is the Fermi energy.
($(\xi_{F}=0,E_{F}=0)$ band touchings will be discussed shortly after.)
Let
$
\mathcal{P}(E)
=
\sum^{q}_{n=1}\,
c_{n}
\left(
E^{2}-E^{2}_{F}
\right)^{n}
- 
\left(\xi^{2} - \xi^{2}_{F}\right)
\,
$
be the Taylor expansion of the characteristic polynomial
Eq.\eqref{eq: E polynomial}
about the band touching point.
The even powers of $E$ in Eq.\eqref{eq: E polynomial}
reflect the spectral particle-hole symmetry, 
and, since $\pm E_{F} \neq 0$ are doubly degenerate roots of the characteristic polynomial, it follows that 
$
\mathcal{P}(E) = (E^{2} - E^{2}_{F})^{2}\,g(E)\,,
$
where $g(E)$ is a polynomial in $E$ of order $2(q-2)$. 
This readily implies the coefficient $c_1 = 0$, leading to the relation in the vicinity of the touching point,
\begin{equation}
\label{eq: xi vs E}
\xi 
\approx
\xi_{F}
+
2 c_2 E^{2}_{F}\xi^{-1}_{F}
\,
(E - E_{F})^{2}
\,,
\quad
\xi_F \neq 0
\,.
\end{equation}
Consequently, the sign of $c_2$ determines 
whether the transition occurs through the quadratic minimum ($\xi_F = \xi_{min} >0$) or maximum ($\xi_F = \xi_{max} >0$) of the Thouless function.
Furthermore, 
upon expanding near the extremal points, i.e.,
$
\xi(\bs{k}) \approx \xi_{min(max)} + \frac{a}{2}\,(\bs{k} - \bs{k}_{min(max)})^{2}
$ 
[with $a > 0 ( < 0)$ being the nonzero curvature at the quadratic minima (maxima)], and substituting onto Eq.\eqref{eq: xi vs E},
we obtain the dispersion 
\begin{equation}
\label{eq: DF dispersion}
E - E_{F} 
=
\pm v^{*}_{F}|\bs{k} - \bs{k}_{min(max)}|
\,,
v^{*}_{F} = (a\xi_{F}/4c_{2}E^{2}_{F})^{1/2}
\,,
\end{equation}
characteristic of a Dirac cone centered at $\bs{k}_{min(max)}$.
{
It can be shown that higher order band touchings are forbidden.
Importantly, we establish below that $\xi$ has $q$ minima and maxima, implying a $q$-component Dirac transition.
}
Figure \ref{fig: band touching} presents two IHCI TPTs for $\phi = \phi_{0}/7$ that confirm the general behavior described in Eq.\eqref{eq: xi vs E} and Eq.\eqref{eq: DF dispersion}.
The considerations above, therefore, uncover a nontrivial link between the classification of critical points and the global properties of the Thouless function, which we now address in detail.

Eq.\eqref{eq: E polynomial} establishes a one-to-one correspondence 
between the zero modes of $\xi$ and band touchings at $E=0$, where
$
E \approx \pm \xi/a^{1/2}_{1}
.$
Then, we directly 
determine from
Eq.\eqref{eq:  xi def}
that 
the band structure with
isotropic hoppings
supports
$2q$ Dirac touchings at $E=0$\cite{Rhim12,Karnaukhov2019,Das2020}
located at
\begin{equation}
\label{eq: isotropic Dirac points}
\bs{K}_{\pm}^{(n)} = 
\left[
\pm \frac{2 \pi}{3 q} + 
\frac{\pi}{q}(2n+q-1)
\,,
\mp\frac{2\pi}{3}
\right)
\,,
\end{equation}
for $n = 0,\cdots, q-1$, and,
furthermore, that these band touchings persist as long as
% \begin{equation}
% \label{eq: center Dirac condition}
% \Big{|}
% \,
% \Big{|}
% \frac{t_1}{t_2}
% \Big{|}^{q}
% -
% 1
% \Big{|}
% \leq
% \Big{|}
% \frac{t_3}{t_2}
% \Big{|}^{q}
% \leq
% \Big{|}
% \,
% \Big{|}
% \frac{t_1}{t_2}
% \Big{|}^{q}
% +
% 1
% \Big{|}
% \,.
% \end{equation}
%%%%
%%%%
%%%%
{
\begin{equation}
\label{eq: center Dirac condition}
\Big{|}|t_i|^{q}- |t_j|^{q}\Big{|} \leq |t_k|^{q} \leq \Big{|}|t_i|^{q} + |t_j|^{q}\Big{|}    
\end{equation}
where $i,j,k$ are identified with any of the distinct values of $1,2,3.$
}
%%%%%
%%%%%
%%%%%
Equation \eqref{eq: center Dirac condition} is the condition for $\xi=0$, which, reproduces the stability of the pair of Dirac cones in graphene bands when $q=1$.\cite{hasegawaPRB2006,WunschNJP2008}
The global properties of the Thouless function lead to a remarkably simple classification of critical points:
\\
\noindent
(1) When the equation \eqref{eq: center Dirac condition} condition holds, $\xi \geq 0$ and there are $2q$ Dirac band touchings at $(\xi=0,E=0)$ {as a consequence of particle-hole symmetry.}
Furthermore, TPTs at nonzero Fermi energy occur through $q$ Dirac band touchings located at
$
\bs{k}^{(n)}_{max} 
= 
[\pi (2n + q-1)/{q},0]
\,,
$
$n = 0, ..., q-1$,
where
$\xi[\bs{k}^{(n)}_{max}] = \xi_{max}$. 
{However, 
$\xi=\xi_{min}=0$  transitions are forbidden at $E \neq 0$ by particle-hole symmetry\cite{supplmat}.}
\\
\noindent
(2) Outside the parameter space \eqref{eq: center Dirac condition}, $\xi > 0$ and the spectrum has a gap at half filling.
The $2q$ zero modes of $\xi$ merge pairwise forming $q$ quadratic minima
at 
one of the saddle points 
$
\bs{M}_1^{(n)} 
= 
\left[\pi (2n+q-1)/q
\,,
-
\pi\,
\right]
\,,
$
$
\bs{M}_2^{(n)} 
= 
\left[ -\pi/q + \pi (2n+q-1)/q 
\,,
0
\right]
$
or
$
\bs{M}_3^{(n)} 
= 
[-\pi/q+\pi (2n+q-1)/q
\,,
-\pi
]
\,,
$
for $n = 0, ..., q-1$.
Then, $E_{F}\neq 0$ critical points are realized by $q$ Dirac band touchings located either at $\xi_{min}$ or $\xi_{max}$.
Taking, for concreteness, \begin{equation}
\label{eq: hoppings}
t_{2} = t_{3} = 1
\,,
\quad
t_1 > 0
\,,
\end{equation}
leads to
case (1) for $0 < t_1 \leq 2^{1/q}$ and case (2) when $t_1 > 2^{1/q}$, where the $q$ degenerate minima of $\xi$ are located at $\bs{k}^{(n)}_{min} = \bs{M}^{(n)}_{2}$,
for $n = 0, ..., q-1$.
The TPTs of Fig.(\ref{fig: band touching})
correspond to case (2) with the hopping parameters Eq.\eqref{eq: hoppings}.
\\
\noindent
(3) The $q$ Dirac fermions at quantum criticality are constrained by the action of magnetic translation, under which $(k_{1},k_{2}) \rightarrow (k_{1} + \frac{2\pi}{q}, k_{2})$,
and they account for
the transfer of Chern number $\Delta\,C = \pm q$ between the bands, 
according to standard parity anomaly considerations.\cite{Redlich1984} We have performed extensive numerical calculations that confirm
the properties (1), (2) and (3).

\begin{figure}[htbp]
\centering
\includegraphics[width = .5\textwidth]{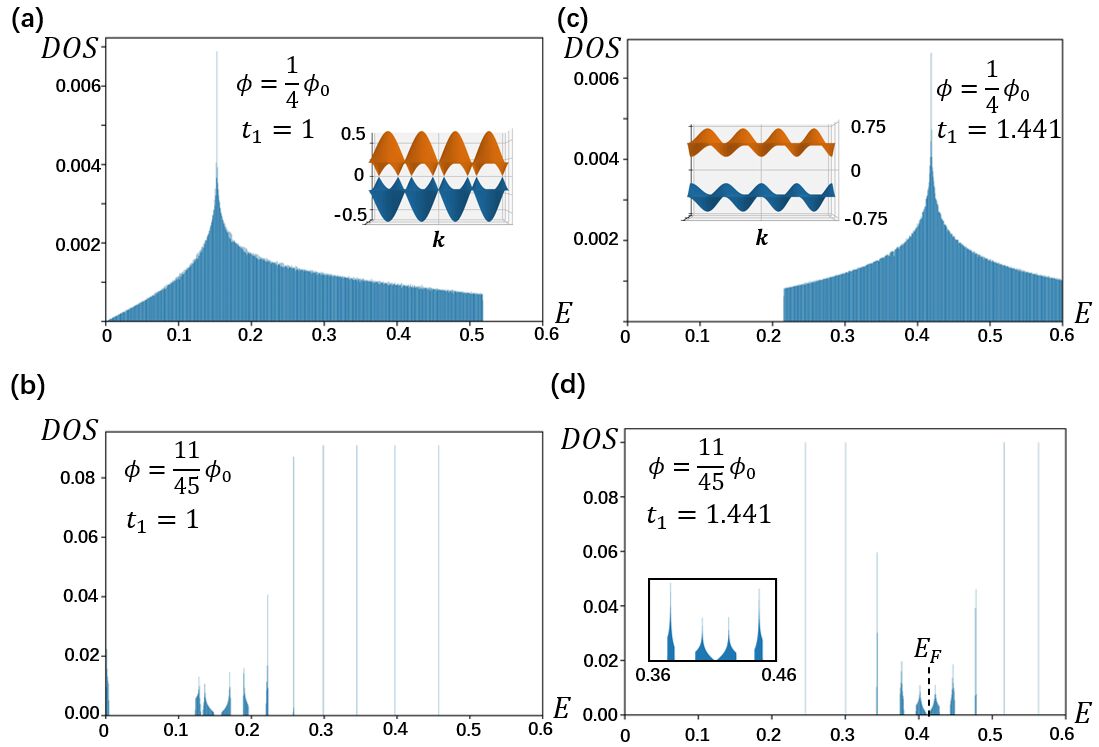}
\caption
{Comparison between the density of states of system A and B.
(a) DOS of the Dirac center band at $\phi_{A}=(1/4)\phi_{0}$ and $(t_1,t_2,t_3) = (1,1,1)$.
Inset: eight gapless DFs with locations given by Eq.\eqref{eq: isotropic Dirac points}.
(b) DOS at $\phi_{B}=(11/45)\phi_{0}$ and $(t_1,t_2,t_3) = (1,1,1)$ reflecting the reconstruction of the Dirac band in (a).
(c) DOS of the Dirac center band at $\phi_{A}=(1/4)\phi_{0}$ and $(t_1,t_2,t_3) = (1.441,1,1)$.
Inset: eight gapped DFs with the gap-opening threshold $t_{1}=2^{1/4}\approx 1.19$.
(d) DOS at $\phi_{B}=(11/45)\phi_{0}$ and $(t_1,t_2,t_3) = (1.441,1,1)$
reflecting the reconstruction of the gapped Dirac band in (c).
Inset shows emergent Dirac fermions at the critical point.
}
\label{fig: DOS}
\end{figure}

Having classified the IHCI critical points, we now address the \textit{mechanism} underlying such phenomena{, which must account for $\Delta C = \pm q$ transitions in a spectrum composed primarily of bands which $C \sim O(1)$. }
Remarkably, 
we argue and numerically demonstrate that {$\Delta C = \pm q$} TPTs occur when Chern bands cross the energy scales associated with the VHS of the DF band close to charge neutrality.
In what follows, we shall demonstrate this 
striking phenomenon using the hopping $t_1$ in Eq.\eqref{eq: hoppings}
as the tuning parameter.

To unearth the connection between VHSs and TPTs, we consider two Hofstadter systems, denoted A and B, with fluxes $\phi_A = p_{A}/q_{A}$ and $\phi_B = p_{B}/q_{B}$
[henceforth we set $h=e=1$ such that $\phi_0 = 1$ and $\phi \sim \phi$ mod (1)].
Furthermore, we impose the conditions
(a) $|(\phi_A - \phi_B)/\phi_{0}| << 1$
and 
(b) 
{$q_{B} \gnapprox
q_{A}$}, 
{which associate the spectrum of B 
with subbands of the A system that arise due to a small residual flux. 
By this construction, the B bands away from the VHS energy $E^{A}_{\textrm{VHS}}$ behave as pseudo-LLs (pLL) of the A system with $C_{\textrm{pLL}} \sim O(1)$. Consequently, we argue, and numerically confirm, that $E^{A}_{\textrm{VHS}}$ provides the natural energy scale supporting nontrivial VHS-Chern bands of B with $C_{\textrm{VHS}} \sim O(q_{B})$.
Therefore, the
dependence of $E^{A}_{\textrm{VHS}}$ on hopping parameters reveals the location of the nontrivial TPTs of B characterized by $\Delta C = \pm q_{B}$.
}

{To gain further insight on the relation between VHSs and TPTs,}
we initially consider system A with $t_i=1$,
which displays $2 q_{A}$ DFs at half filling
with $E^{A}_{Dirac}(\bs{k})
\approx \xi_{A}(t_i=1;\bs{k} - \bs{K}_{\pm})/a_{1}^{1/2}$; see Eq.\eqref{eq: isotropic Dirac points}.
Due to particle-hole symmetry, we focus on $E\geq 0$ bands.
General considerations 
give the Dirac-like 
density of states (DOS) $D_{A} \propto E$ near charge neutrality,
which is cut off by the VHS energy $E^{A}_{\textrm{VHS}}$ 
that distinguishes the electronlike states from the holelike states. 
Figure \ref{fig: DOS}(a) displays the DOS of this band for $\phi_A = 1/4$, which supports eight Dirac fermions and has
$E^{A}_{\textrm{VHS}} \approx 0.15$.
Notice that, compared to the graphene bands\cite{graphene-RMP}, the magnetic field pushes the VHS substantially closer to charge neutrality due to the splitting of the spectrum into $2q_{A}$ bands.
Furthermore, conditions (a) and (b) ensure the spectrum of B near half filling can be understood as the response of the DF band of A to a weak ``residual" magnetic field, which is expected to give rise to relativisticlike (nonrelativistic-like) LLs for 
$
E 
\lesssim 
(\gtrsim) 
E^{A}_{\textrm{VHS}}
\,.
$
However, 
the B bands close to
$E^{A}_{\textrm{VHS}}$
deviate substantially from the LL behavior{, confirming the behavior described in the paragraph above.}
This is illustrated 
in Fig.\ref{fig: DOS}(b) where the said bands of the $\phi_{B} = 11/45$ system show more pronounced bandwidths and narrower gaps.
%
% {Because $E^{A}_{\textrm{VHS}}$ changes with the hopping parameters,
% %
% the change in
% $t_1$ 
% away from the isotropic point
% steers the ``VHS-Chern bands" of B, 
% as shown in Figs.(\ref{fig: DOS}c) and (\ref{fig: DOS}d). This process reveals
% a sequence of TPTs
% characterized by $\Delta C  = \pm 45$.
% %
% Fig.(\ref{fig: phase transitions}) confirms the mechanism discussed above that
% the bands participating in the TPTs lie in the energy scale $E^{A}_{\textrm{VHS}}$, represented by the solid green and purple lines.
% }
%

{To understand how $E^{A}_{\textrm{VHS}}$ tracks 
the TPTs of the B system, we study 
the the dependence of Thouless function on the hopping parameters.}
The property
$E_{\alpha}(\bs{k}) = E_{\alpha}[\xi(\bs{k})]$ 
establishes that the VHSs of the Chern bands are located on the saddles of $\xi$.
Direct calculation
shows that $\xi$ is degenerate on all the saddle points $\bs{M}^{(n)}_{1,2,3}$ when $t_1=1$ and, furthermore, that the degeneracy is partially broken for $t_1 \neq 1$.\cite{supplmat}
For $1 < t_1 < 2^{1/q}$ [case (1) above], the VHS splits into a large peak at $E^{A}_{\textrm{VHS,1}}\equiv E^{A}[\bs{M}^{(n)}_{1}]$ 
and a small peak at $E^{A}_{\textrm{VHS,2}}\equiv E^{A}(\bs{M}^{(n)}_{2})$.
The latter disappears in the lower band edge, 
for $t_1 > 2^{1/q}$ [case (2) above],
where an energy gap forms [Fig. \ref{fig: DOS}(c)].
Moreover, Fig. \ref{fig: DOS}(d) (see inset) displays the onset of a TPT 
as the result of the VHS-Chern bands being steered by the $E^{A}_{\textrm{VHS,1}}$ energy scale.

\begin{figure}[htbp]
\centering
\includegraphics[width = .5\textwidth]{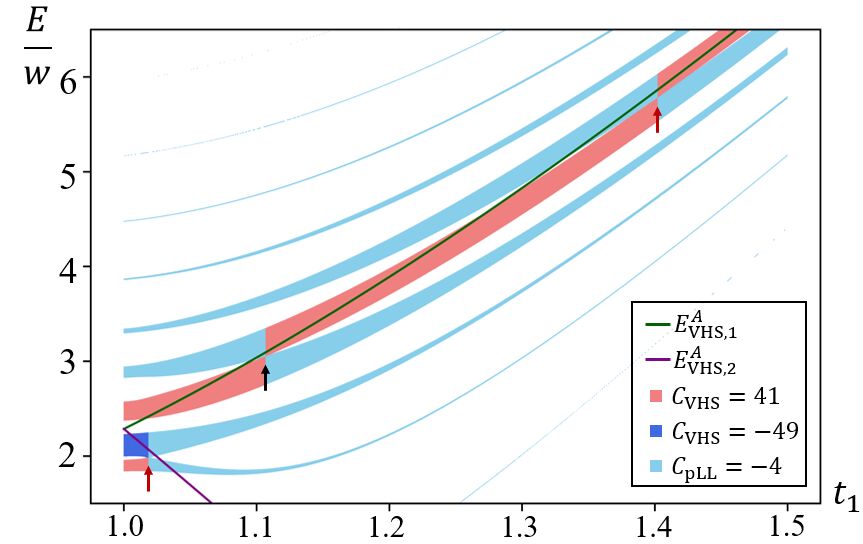}
\caption{
TPTs of system B ($\phi_{B} = 11/45$) steered by the VHSs of system A ($\phi_{A} = 1/4$).
{Eleven B bands form near charge neutrality by splitting of the Dirac band of A in response to a flux deviation $\delta \phi = -1/180$, where nine of these bands are shown.}
All energies are rescaled by the average band separation $w$ of the B system with $t_1=1$.
{The Chern numbers of the bands are indicated by color coding.}  
$E^{A}_{\textrm{VHS,1}}$ and $E^{A}_{\textrm{VHS,2}}$ are represented, respectively, by solid green and purple lines.
%
%\sout{Dashed green and purple lines represent  $E^{A}_{Dirac} = \xi/\sqrt{a_1}$
%evaluated, respectively, at the saddle points $\bs{M}_{1}$ and $\bs{M}_{2}$ of $\xi$.}
%
%
The composite fermion {(IHCI)} TPTs at $n=47/90$ and $n=50/90$ {($n = 49/90$)} are marked by vertical red {(black)} arrows. 
}
\label{fig: phase transitions}
\end{figure}

{The striking relationship between VHS and TPTs is shown in 
Fig. \ref{fig: phase transitions}, where the bands of the $\phi_B = p_{B}/q_{B} = 11/45$ system near charge neutrality are plotted in the interval $t_1 \geq 1$.
These bands originate as subbands of the
$\phi_A =  p_{A}/q_{A} = 1/4$ Dirac band in response to a small flux deviation $\delta \phi = -1/180$, as per conditions (a) and (b).
We observe that the B bands formed near the band edges of system A behave as pLLs with vanishing bandwidth and $C_{\textrm{pLL}}=-4$, while the VHS-Chern bands carrying $C_{\textrm{VHS}} \sim O(q_{B})$ form 
in the vicinity of $E^{A}_{\textrm{VHS}}$.
Because $E^{A}_{\textrm{VHS}}$ changes with the hopping parameters,
the change in
$t_1$ 
away from the isotropic point
steers the VHS-Chern bands of B along the solid green ($E^{A}_{\textrm{VHS},1}$) and purple ($E^{A}_{\textrm{VHS},2}$) lines.
This VHS steering mechanism reveals
a sequence of TPTs (up arrows)
characterized by $\Delta C = \pm 45$, 
with $45$ emerging DFs located at the
extremum points of the Thouless function of the system B, confirming the general properties (1), (2) and (3).
%
%\sout{Moreover, we have found that the VHS steering mechanism is ubiquitous and applies to Chern bands %associated with magnetic fluxes other than the ones discussed here.\cite{supplmat}}
}
{For results on other flux states, see the Supplemental Material\cite{supplmat}}.

\noindent
\textbf{FHCI transitions -- } 
Our analysis can be further extended to describe
FHCI transitions tuned by the hopping parameters in partially filled Chern bands.
via the standard representation of an FHCI with Hall conductance
$\sigma_{xy}(C) = C/(2C+1)$
in terms of a composite fermion system \cite{Jain1989,Lopez1991,HLR}
in an IHCI with $\sigma^{CF}_{xy} = C$\cite{Kol1993,Moller2015}, which is
subject to a mean field
residual flux 
\begin{equation}
\label{eq: flux attachment}
\phi_{CF}
=
\phi
-
\phi_{CS}
\,,
\end{equation}
where $\phi = B\,{\frac{\sqrt{3}}{2}}a^{2}$ and $\phi_{CS}=4n$ (the factor of $4$ accounts for two attached flux quanta and two sites per unit cell) are, respectively, the intracell fluxes due to the external magnetic field and the Chern-Simons gauge field at lattice filling $n$, for $0 \leq n \leq 1$.
Then, a TPT at fixed $B$ and $n$ between FHCIs with
$\sigma_{xy}(C_1) = C_{1}/(2C_{1}+1)$
and
$\sigma_{xy}(C_2) = C_{2}/(2C_{2}+1)$
can be effectively described by a $C_{1} \rightarrow C_{2}$
composite fermion transition 
subject to the constraint $|C_2 - C_1| = q_{cf}$ (recall  property (3)), where $\phi_{CF} = p_{cf}/q_{cf}$ is the flux of the composite fermion state.
Furthermore, the relationship Eq.\eqref{eq: flux attachment} between $B$ and $n$ allows the identification of candidate TPTs between Abelian FHCI states. 
In closing we present two such FHCI transitions realized when
$\phi_{CF} = 11/45$, which are shown by vertical red arrows in Fig. \ref{fig: phase transitions}.
The first TPT is observed at 
($t_1 \approx 1.02, n = 47/9, \phi = 1/3$)
and represents a transition between FHCIs with
$\sigma_{xy}(37) = 37/75$
and
$\sigma_{xy}(-8) = 8/15$.
On the second transition at ($t_1 \approx 1.44, n = 50/90 = 5/9, \phi = 7/15$),
the Hall conductance jumps from
$\sigma_{xy}(25) = 25/51$ to
$\sigma_{xy}(-20) = 20/39$.
We point the reader to the Supplemental Material \cite{supplmat} for another example of FHCI transition.

In summary, 
we have proposed an
analytical framework
to classify multiflavor Dirac fermion critical points
describing hopping-tuned TPTs of integer and fractional Hofstadter-Chern insulators in honeycomb superlattices.
Our classification sets firm constraints on the number of Dirac flavors as well as their momentum space distribution in terms of the hopping parameters, the magnetic flux per unit cell and the electron density.
Such critical points realize large transfers of Chern number across the TPT, which can be detected via conductivity measurements.
We have identified a series of TPTs that can be explained by the nontrivial response of Chern bands to VHSs near charge neutrality.
These results, which were derived from the identification of global properties of the Chern bands, lead to a new understanding of quantum critical phenomena resulting from the interplay of magnetic fields and VHSs.
This work opens many interesting directions to study quantum critical phenomena in superlattices.
Besides nanopatterned graphene superlattices~\cite{Gibertini2009,Singha2011,Soibel_1996,N_dvorn_k_2012,Wang2016,Wang2018} that served as a motivation for this work,
van der Waals heterostructures in external magnetic field\cite{Dean2013,Ponomarenko2013,Hunt2013,Spanton2018} 
provide promising platforms to realize topological quantum criticality via strain induced tuning of the effective hopping parameters.
Also, the interplay of magnetic fields and higher order VHSs\cite{efremov2019,Yuan2019}
can potentially provide even richer critical phenomena.
We leave these open questions to future work.

\begin{acknowledgements}
We thank Claudio Chamon, Ankur Das, Ribhu Kaul, Ganpathy Murthy and Raman Sohal for useful discussions.
Part of this work was performed at the Aspen Center for Physics, which is supported by National Science Foundation Grant No. PHY-1607611.
L. H. S. is supported by a faculty startup at Emory University.
\end{acknowledgements}

\noindent
\textit{Note added in proof}--Recently, we became aware of a related work, Ref.\cite{Bernevig2020}, which studies quantum phase transitions in Hofstadter bands by tuning the magnetic flux per unit cell.

%%%%%%%%%%%%%%%%%%%%%%%%%%%%%%%%
\bibliographystyle{apsrev4-1}
\bibliography{arxivupdate.bib}

\begin{thebibliography}{51}%
\makeatletter
\providecommand \@ifxundefined [1]{%
 \@ifx{#1\undefined}
}%
\providecommand \@ifnum [1]{%
 \ifnum #1\expandafter \@firstoftwo
 \else \expandafter \@secondoftwo
 \fi
}%
\providecommand \@ifx [1]{%
 \ifx #1\expandafter \@firstoftwo
 \else \expandafter \@secondoftwo
 \fi
}%
\providecommand \natexlab [1]{#1}%
\providecommand \enquote  [1]{``#1''}%
\providecommand \bibnamefont  [1]{#1}%
\providecommand \bibfnamefont [1]{#1}%
\providecommand \citenamefont [1]{#1}%
\providecommand \href@noop [0]{\@secondoftwo}%
\providecommand \href [0]{\begingroup \@sanitize@url \@href}%
\providecommand \@href[1]{\@@startlink{#1}\@@href}%
\providecommand \@@href[1]{\endgroup#1\@@endlink}%
\providecommand \@sanitize@url [0]{\catcode `\\12\catcode `\$12\catcode
  `\&12\catcode `\#12\catcode `\^12\catcode `\_12\catcode `\%12\relax}%
\providecommand \@@startlink[1]{}%
\providecommand \@@endlink[0]{}%
\providecommand \url  [0]{\begingroup\@sanitize@url \@url }%
\providecommand \@url [1]{\endgroup\@href {#1}{\urlprefix }}%
\providecommand \urlprefix  [0]{URL }%
\providecommand \Eprint [0]{\href }%
\providecommand \doibase [0]{http://dx.doi.org/}%
\providecommand \selectlanguage [0]{\@gobble}%
\providecommand \bibinfo  [0]{\@secondoftwo}%
\providecommand \bibfield  [0]{\@secondoftwo}%
\providecommand \translation [1]{[#1]}%
\providecommand \BibitemOpen [0]{}%
\providecommand \bibitemStop [0]{}%
\providecommand \bibitemNoStop [0]{.\EOS\space}%
\providecommand \EOS [0]{\spacefactor3000\relax}%
\providecommand \BibitemShut  [1]{\csname bibitem#1\endcsname}%
\let\auto@bib@innerbib\@empty
%</preamble>
\bibitem [{\citenamefont {Thouless}\ \emph {et~al.}(1982)\citenamefont
  {Thouless}, \citenamefont {Kohmoto}, \citenamefont {Nightingale},\ and\
  \citenamefont {den Nijs}}]{TKNN1982}%
  \BibitemOpen
  \bibfield  {author} {\bibinfo {author} {\bibfnamefont {D.~J.}\ \bibnamefont
  {Thouless}}, \bibinfo {author} {\bibfnamefont {M.}~\bibnamefont {Kohmoto}},
  \bibinfo {author} {\bibfnamefont {M.~P.}\ \bibnamefont {Nightingale}}, \ and\
  \bibinfo {author} {\bibfnamefont {M.}~\bibnamefont {den Nijs}},\ }\href
  {\doibase 10.1103/PhysRevLett.49.405} {\bibfield  {journal} {\bibinfo
  {journal} {Phys. Rev. Lett.}\ }\textbf {\bibinfo {volume} {49}},\ \bibinfo
  {pages} {405} (\bibinfo {year} {1982})}\BibitemShut {NoStop}%
\bibitem [{\citenamefont {Haldane}(1988)}]{Haldane1988}%
  \BibitemOpen
  \bibfield  {author} {\bibinfo {author} {\bibfnamefont {F.~D.~M.}\
  \bibnamefont {Haldane}},\ }\href {\doibase 10.1103/PhysRevLett.61.2015}
  {\bibfield  {journal} {\bibinfo  {journal} {Phys. Rev. Lett.}\ }\textbf
  {\bibinfo {volume} {61}},\ \bibinfo {pages} {2015} (\bibinfo {year}
  {1988})}\BibitemShut {NoStop}%
\bibitem [{\citenamefont {Hofstadter}(1976)}]{Hofstadter1976}%
  \BibitemOpen
  \bibfield  {author} {\bibinfo {author} {\bibfnamefont {D.~R.}\ \bibnamefont
  {Hofstadter}},\ }\href {\doibase 10.1103/PhysRevB.14.2239} {\bibfield
  {journal} {\bibinfo  {journal} {Phys. Rev. B}\ }\textbf {\bibinfo {volume}
  {14}},\ \bibinfo {pages} {2239} (\bibinfo {year} {1976})}\BibitemShut
  {NoStop}%
\bibitem [{\citenamefont {Neupert}\ \emph {et~al.}(2011)\citenamefont
  {Neupert}, \citenamefont {Santos}, \citenamefont {Chamon},\ and\
  \citenamefont {Mudry}}]{Neupert-2011}%
  \BibitemOpen
  \bibfield  {author} {\bibinfo {author} {\bibfnamefont {T.}~\bibnamefont
  {Neupert}}, \bibinfo {author} {\bibfnamefont {L.}~\bibnamefont {Santos}},
  \bibinfo {author} {\bibfnamefont {C.}~\bibnamefont {Chamon}}, \ and\ \bibinfo
  {author} {\bibfnamefont {C.}~\bibnamefont {Mudry}},\ }\href {\doibase
  10.1103/PhysRevLett.106.236804} {\bibfield  {journal} {\bibinfo  {journal}
  {Phys. Rev. Lett.}\ }\textbf {\bibinfo {volume} {106}},\ \bibinfo {pages}
  {236804} (\bibinfo {year} {2011})}\BibitemShut {NoStop}%
\bibitem [{\citenamefont {Sheng}\ \emph {et~al.}(2011)\citenamefont {Sheng},
  \citenamefont {Gu}, \citenamefont {Sun},\ and\ \citenamefont
  {Sheng}}]{Sheng-2011}%
  \BibitemOpen
  \bibfield  {author} {\bibinfo {author} {\bibfnamefont {D.~N.}\ \bibnamefont
  {Sheng}}, \bibinfo {author} {\bibfnamefont {Z.-C.}\ \bibnamefont {Gu}},
  \bibinfo {author} {\bibfnamefont {K.}~\bibnamefont {Sun}}, \ and\ \bibinfo
  {author} {\bibfnamefont {L.}~\bibnamefont {Sheng}},\ }\href {\doibase
  10.1038/ncomms1380} {\bibfield  {journal} {\bibinfo  {journal} {Nature
  Communications}\ }\textbf {\bibinfo {volume} {2}},\ \bibinfo {pages} {389 EP
  } (\bibinfo {year} {2011})}\BibitemShut {NoStop}%
\bibitem [{\citenamefont {Tang}\ \emph {et~al.}(2011)\citenamefont {Tang},
  \citenamefont {Mei},\ and\ \citenamefont {Wen}}]{Tang-2011}%
  \BibitemOpen
  \bibfield  {author} {\bibinfo {author} {\bibfnamefont {E.}~\bibnamefont
  {Tang}}, \bibinfo {author} {\bibfnamefont {J.-W.}\ \bibnamefont {Mei}}, \
  and\ \bibinfo {author} {\bibfnamefont {X.-G.}\ \bibnamefont {Wen}},\ }\href
  {\doibase 10.1103/PhysRevLett.106.236802} {\bibfield  {journal} {\bibinfo
  {journal} {Phys. Rev. Lett.}\ }\textbf {\bibinfo {volume} {106}},\ \bibinfo
  {pages} {236802} (\bibinfo {year} {2011})}\BibitemShut {NoStop}%
\bibitem [{\citenamefont {Sun}\ \emph {et~al.}(2011)\citenamefont {Sun},
  \citenamefont {Gu}, \citenamefont {Katsura},\ and\ \citenamefont
  {Das~Sarma}}]{Sun-2011}%
  \BibitemOpen
  \bibfield  {author} {\bibinfo {author} {\bibfnamefont {K.}~\bibnamefont
  {Sun}}, \bibinfo {author} {\bibfnamefont {Z.}~\bibnamefont {Gu}}, \bibinfo
  {author} {\bibfnamefont {H.}~\bibnamefont {Katsura}}, \ and\ \bibinfo
  {author} {\bibfnamefont {S.}~\bibnamefont {Das~Sarma}},\ }\href {\doibase
  10.1103/PhysRevLett.106.236803} {\bibfield  {journal} {\bibinfo  {journal}
  {Phys. Rev. Lett.}\ }\textbf {\bibinfo {volume} {106}},\ \bibinfo {pages}
  {236803} (\bibinfo {year} {2011})}\BibitemShut {NoStop}%
\bibitem [{\citenamefont {Regnault}\ and\ \citenamefont
  {Bernevig}(2011)}]{Regnault2011}%
  \BibitemOpen
  \bibfield  {author} {\bibinfo {author} {\bibfnamefont {N.}~\bibnamefont
  {Regnault}}\ and\ \bibinfo {author} {\bibfnamefont {B.~A.}\ \bibnamefont
  {Bernevig}},\ }\href {\doibase 10.1103/PhysRevX.1.021014} {\bibfield
  {journal} {\bibinfo  {journal} {Phys. Rev. X}\ }\textbf {\bibinfo {volume}
  {1}},\ \bibinfo {pages} {021014} (\bibinfo {year} {2011})}\BibitemShut
  {NoStop}%
\bibitem [{\citenamefont {Dean}\ \emph {et~al.}(2013)\citenamefont {Dean},
  \citenamefont {Wang}, \citenamefont {Maher}, \citenamefont {Forsythe},
  \citenamefont {Ghahari}, \citenamefont {Gao}, \citenamefont {Katoch},
  \citenamefont {Ishigami}, \citenamefont {Moon}, \citenamefont {Koshino},
  \citenamefont {Taniguchi}, \citenamefont {Watanabe}, \citenamefont {Shepard},
  \citenamefont {Hone},\ and\ \citenamefont {Kim}}]{Dean2013}%
  \BibitemOpen
  \bibfield  {author} {\bibinfo {author} {\bibfnamefont {C.~R.}\ \bibnamefont
  {Dean}}, \bibinfo {author} {\bibfnamefont {L.}~\bibnamefont {Wang}}, \bibinfo
  {author} {\bibfnamefont {P.}~\bibnamefont {Maher}}, \bibinfo {author}
  {\bibfnamefont {C.}~\bibnamefont {Forsythe}}, \bibinfo {author}
  {\bibfnamefont {F.}~\bibnamefont {Ghahari}}, \bibinfo {author} {\bibfnamefont
  {Y.}~\bibnamefont {Gao}}, \bibinfo {author} {\bibfnamefont {J.}~\bibnamefont
  {Katoch}}, \bibinfo {author} {\bibfnamefont {M.}~\bibnamefont {Ishigami}},
  \bibinfo {author} {\bibfnamefont {P.}~\bibnamefont {Moon}}, \bibinfo {author}
  {\bibfnamefont {M.}~\bibnamefont {Koshino}}, \bibinfo {author} {\bibfnamefont
  {T.}~\bibnamefont {Taniguchi}}, \bibinfo {author} {\bibfnamefont
  {K.}~\bibnamefont {Watanabe}}, \bibinfo {author} {\bibfnamefont {K.~L.}\
  \bibnamefont {Shepard}}, \bibinfo {author} {\bibfnamefont {J.}~\bibnamefont
  {Hone}}, \ and\ \bibinfo {author} {\bibfnamefont {P.}~\bibnamefont {Kim}},\
  }\href {https://doi.org/10.1038/nature12186} {\bibfield  {journal} {\bibinfo
  {journal} {Nature}\ }\textbf {\bibinfo {volume} {497}},\ \bibinfo {pages}
  {598 EP } (\bibinfo {year} {2013})}\BibitemShut {NoStop}%
\bibitem [{\citenamefont {Ponomarenko}\ \emph {et~al.}(2013)\citenamefont
  {Ponomarenko}, \citenamefont {Gorbachev}, \citenamefont {Yu}, \citenamefont
  {Elias}, \citenamefont {Jalil}, \citenamefont {Patel}, \citenamefont
  {Mishchenko}, \citenamefont {Mayorov}, \citenamefont {Woods}, \citenamefont
  {Wallbank}, \citenamefont {Mucha-Kruczynski}, \citenamefont {Piot},
  \citenamefont {Potemski}, \citenamefont {Grigorieva}, \citenamefont
  {Novoselov}, \citenamefont {Guinea}, \citenamefont {Fal'ko},\ and\
  \citenamefont {Geim}}]{Ponomarenko2013}%
  \BibitemOpen
  \bibfield  {author} {\bibinfo {author} {\bibfnamefont {L.~A.}\ \bibnamefont
  {Ponomarenko}}, \bibinfo {author} {\bibfnamefont {R.~V.}\ \bibnamefont
  {Gorbachev}}, \bibinfo {author} {\bibfnamefont {G.~L.}\ \bibnamefont {Yu}},
  \bibinfo {author} {\bibfnamefont {D.~C.}\ \bibnamefont {Elias}}, \bibinfo
  {author} {\bibfnamefont {R.}~\bibnamefont {Jalil}}, \bibinfo {author}
  {\bibfnamefont {A.~A.}\ \bibnamefont {Patel}}, \bibinfo {author}
  {\bibfnamefont {A.}~\bibnamefont {Mishchenko}}, \bibinfo {author}
  {\bibfnamefont {A.~S.}\ \bibnamefont {Mayorov}}, \bibinfo {author}
  {\bibfnamefont {C.~R.}\ \bibnamefont {Woods}}, \bibinfo {author}
  {\bibfnamefont {J.~R.}\ \bibnamefont {Wallbank}}, \bibinfo {author}
  {\bibfnamefont {M.}~\bibnamefont {Mucha-Kruczynski}}, \bibinfo {author}
  {\bibfnamefont {B.~A.}\ \bibnamefont {Piot}}, \bibinfo {author}
  {\bibfnamefont {M.}~\bibnamefont {Potemski}}, \bibinfo {author}
  {\bibfnamefont {I.~V.}\ \bibnamefont {Grigorieva}}, \bibinfo {author}
  {\bibfnamefont {K.~S.}\ \bibnamefont {Novoselov}}, \bibinfo {author}
  {\bibfnamefont {F.}~\bibnamefont {Guinea}}, \bibinfo {author} {\bibfnamefont
  {V.~I.}\ \bibnamefont {Fal'ko}}, \ and\ \bibinfo {author} {\bibfnamefont
  {A.~K.}\ \bibnamefont {Geim}},\ }\href {https://doi.org/10.1038/nature12187}
  {\bibfield  {journal} {\bibinfo  {journal} {Nature}\ }\textbf {\bibinfo
  {volume} {497}},\ \bibinfo {pages} {594 EP } (\bibinfo {year}
  {2013})}\BibitemShut {NoStop}%
\bibitem [{\citenamefont {Hunt}\ \emph {et~al.}(2013)\citenamefont {Hunt},
  \citenamefont {Sanchez-Yamagishi}, \citenamefont {Young}, \citenamefont
  {Yankowitz}, \citenamefont {LeRoy}, \citenamefont {Watanabe}, \citenamefont
  {Taniguchi}, \citenamefont {Moon}, \citenamefont {Koshino}, \citenamefont
  {Jarillo-Herrero},\ and\ \citenamefont {Ashoori}}]{Hunt2013}%
  \BibitemOpen
  \bibfield  {author} {\bibinfo {author} {\bibfnamefont {B.}~\bibnamefont
  {Hunt}}, \bibinfo {author} {\bibfnamefont {J.~D.}\ \bibnamefont
  {Sanchez-Yamagishi}}, \bibinfo {author} {\bibfnamefont {A.~F.}\ \bibnamefont
  {Young}}, \bibinfo {author} {\bibfnamefont {M.}~\bibnamefont {Yankowitz}},
  \bibinfo {author} {\bibfnamefont {B.~J.}\ \bibnamefont {LeRoy}}, \bibinfo
  {author} {\bibfnamefont {K.}~\bibnamefont {Watanabe}}, \bibinfo {author}
  {\bibfnamefont {T.}~\bibnamefont {Taniguchi}}, \bibinfo {author}
  {\bibfnamefont {P.}~\bibnamefont {Moon}}, \bibinfo {author} {\bibfnamefont
  {M.}~\bibnamefont {Koshino}}, \bibinfo {author} {\bibfnamefont
  {P.}~\bibnamefont {Jarillo-Herrero}}, \ and\ \bibinfo {author} {\bibfnamefont
  {R.~C.}\ \bibnamefont {Ashoori}},\ }\href {\doibase 10.1126/science.1237240}
  {\bibfield  {journal} {\bibinfo  {journal} {Science}\ }\textbf {\bibinfo
  {volume} {340}},\ \bibinfo {pages} {1427} (\bibinfo {year}
  {2013})}\BibitemShut {NoStop}%
\bibitem [{\citenamefont {Forsythe}\ \emph {et~al.}(2018)\citenamefont
  {Forsythe}, \citenamefont {Zhou}, \citenamefont {Watanabe}, \citenamefont
  {Taniguchi}, \citenamefont {Pasupathy}, \citenamefont {Moon}, \citenamefont
  {Koshino}, \citenamefont {Kim},\ and\ \citenamefont
  {Dean}}]{forsythe-NatureNanoTech2018}%
  \BibitemOpen
  \bibfield  {author} {\bibinfo {author} {\bibfnamefont {C.}~\bibnamefont
  {Forsythe}}, \bibinfo {author} {\bibfnamefont {X.}~\bibnamefont {Zhou}},
  \bibinfo {author} {\bibfnamefont {K.}~\bibnamefont {Watanabe}}, \bibinfo
  {author} {\bibfnamefont {T.}~\bibnamefont {Taniguchi}}, \bibinfo {author}
  {\bibfnamefont {A.}~\bibnamefont {Pasupathy}}, \bibinfo {author}
  {\bibfnamefont {P.}~\bibnamefont {Moon}}, \bibinfo {author} {\bibfnamefont
  {M.}~\bibnamefont {Koshino}}, \bibinfo {author} {\bibfnamefont
  {P.}~\bibnamefont {Kim}}, \ and\ \bibinfo {author} {\bibfnamefont {C.~R.}\
  \bibnamefont {Dean}},\ }\href@noop {} {\bibfield  {journal} {\bibinfo
  {journal} {Nature nanotechnology}\ }\textbf {\bibinfo {volume} {13}},\
  \bibinfo {pages} {566} (\bibinfo {year} {2018})}\BibitemShut {NoStop}%
\bibitem [{\citenamefont {Spanton}\ \emph {et~al.}(2018)\citenamefont
  {Spanton}, \citenamefont {Zibrov}, \citenamefont {Zhou}, \citenamefont
  {Taniguchi}, \citenamefont {Watanabe}, \citenamefont {Zaletel},\ and\
  \citenamefont {Young}}]{Spanton2018}%
  \BibitemOpen
  \bibfield  {author} {\bibinfo {author} {\bibfnamefont {E.~M.}\ \bibnamefont
  {Spanton}}, \bibinfo {author} {\bibfnamefont {A.~A.}\ \bibnamefont {Zibrov}},
  \bibinfo {author} {\bibfnamefont {H.}~\bibnamefont {Zhou}}, \bibinfo {author}
  {\bibfnamefont {T.}~\bibnamefont {Taniguchi}}, \bibinfo {author}
  {\bibfnamefont {K.}~\bibnamefont {Watanabe}}, \bibinfo {author}
  {\bibfnamefont {M.~P.}\ \bibnamefont {Zaletel}}, \ and\ \bibinfo {author}
  {\bibfnamefont {A.~F.}\ \bibnamefont {Young}},\ }\href {\doibase
  10.1126/science.aan8458} {\bibfield  {journal} {\bibinfo  {journal}
  {Science}\ }\textbf {\bibinfo {volume} {360}},\ \bibinfo {pages} {62}
  (\bibinfo {year} {2018})}\BibitemShut {NoStop}%
\bibitem [{\citenamefont {Liu}\ \emph {et~al.}(2012)\citenamefont {Liu},
  \citenamefont {Bergholtz}, \citenamefont {Fan},\ and\ \citenamefont
  {L\"auchli}}]{Liu-PRL2012}%
  \BibitemOpen
  \bibfield  {author} {\bibinfo {author} {\bibfnamefont {Z.}~\bibnamefont
  {Liu}}, \bibinfo {author} {\bibfnamefont {E.~J.}\ \bibnamefont {Bergholtz}},
  \bibinfo {author} {\bibfnamefont {H.}~\bibnamefont {Fan}}, \ and\ \bibinfo
  {author} {\bibfnamefont {A.~M.}\ \bibnamefont {L\"auchli}},\ }\href {\doibase
  10.1103/PhysRevLett.109.186805} {\bibfield  {journal} {\bibinfo  {journal}
  {Phys. Rev. Lett.}\ }\textbf {\bibinfo {volume} {109}},\ \bibinfo {pages}
  {186805} (\bibinfo {year} {2012})}\BibitemShut {NoStop}%
\bibitem [{\citenamefont {Wu}\ \emph {et~al.}(2012)\citenamefont {Wu},
  \citenamefont {Bernevig},\ and\ \citenamefont {Regnault}}]{Wu-PRB2012}%
  \BibitemOpen
  \bibfield  {author} {\bibinfo {author} {\bibfnamefont {Y.-L.}\ \bibnamefont
  {Wu}}, \bibinfo {author} {\bibfnamefont {B.~A.}\ \bibnamefont {Bernevig}}, \
  and\ \bibinfo {author} {\bibfnamefont {N.}~\bibnamefont {Regnault}},\ }\href
  {\doibase 10.1103/PhysRevB.85.075116} {\bibfield  {journal} {\bibinfo
  {journal} {Phys. Rev. B}\ }\textbf {\bibinfo {volume} {85}},\ \bibinfo
  {pages} {075116} (\bibinfo {year} {2012})}\BibitemShut {NoStop}%
\bibitem [{\citenamefont {L\"auchli}\ \emph {et~al.}(2013)\citenamefont
  {L\"auchli}, \citenamefont {Liu}, \citenamefont {Bergholtz},\ and\
  \citenamefont {Moessner}}]{Lauchli-PRL2013}%
  \BibitemOpen
  \bibfield  {author} {\bibinfo {author} {\bibfnamefont {A.~M.}\ \bibnamefont
  {L\"auchli}}, \bibinfo {author} {\bibfnamefont {Z.}~\bibnamefont {Liu}},
  \bibinfo {author} {\bibfnamefont {E.~J.}\ \bibnamefont {Bergholtz}}, \ and\
  \bibinfo {author} {\bibfnamefont {R.}~\bibnamefont {Moessner}},\ }\href
  {\doibase 10.1103/PhysRevLett.111.126802} {\bibfield  {journal} {\bibinfo
  {journal} {Phys. Rev. Lett.}\ }\textbf {\bibinfo {volume} {111}},\ \bibinfo
  {pages} {126802} (\bibinfo {year} {2013})}\BibitemShut {NoStop}%
\bibitem [{\citenamefont {Jain}(1989)}]{Jain1989}%
  \BibitemOpen
  \bibfield  {author} {\bibinfo {author} {\bibfnamefont {J.~K.}\ \bibnamefont
  {Jain}},\ }\href {\doibase https://doi.org/10.1103/PhysRevLett.63.199}
  {\bibfield  {journal} {\bibinfo  {journal} {Phys. Rev. Lett.}\ }\textbf
  {\bibinfo {volume} {63}},\ \bibinfo {pages} {199} (\bibinfo {year}
  {1989})}\BibitemShut {NoStop}%
\bibitem [{\citenamefont {L{\'o}pez}\ and\ \citenamefont
  {Fradkin}(1991)}]{Lopez1991}%
  \BibitemOpen
  \bibfield  {author} {\bibinfo {author} {\bibfnamefont {A.}~\bibnamefont
  {L{\'o}pez}}\ and\ \bibinfo {author} {\bibfnamefont {E.}~\bibnamefont
  {Fradkin}},\ }\href {\doibase https://doi.org/10.1103/PhysRevB.44.5246}
  {\bibfield  {journal} {\bibinfo  {journal} {Phys. Rev. B}\ }\textbf {\bibinfo
  {volume} {44}},\ \bibinfo {pages} {5246} (\bibinfo {year}
  {1991})}\BibitemShut {NoStop}%
\bibitem [{\citenamefont {Kol}\ and\ \citenamefont {Read}(1993)}]{Kol1993}%
  \BibitemOpen
  \bibfield  {author} {\bibinfo {author} {\bibfnamefont {A.}~\bibnamefont
  {Kol}}\ and\ \bibinfo {author} {\bibfnamefont {N.}~\bibnamefont {Read}},\
  }\href {\doibase 10.1103/PhysRevB.48.8890} {\bibfield  {journal} {\bibinfo
  {journal} {Phys. Rev. B}\ }\textbf {\bibinfo {volume} {48}},\ \bibinfo
  {pages} {8890} (\bibinfo {year} {1993})}\BibitemShut {NoStop}%
\bibitem [{\citenamefont {M\"oller}\ and\ \citenamefont
  {Cooper}(2015)}]{Moller2015}%
  \BibitemOpen
  \bibfield  {author} {\bibinfo {author} {\bibfnamefont {G.}~\bibnamefont
  {M\"oller}}\ and\ \bibinfo {author} {\bibfnamefont {N.~R.}\ \bibnamefont
  {Cooper}},\ }\href {\doibase 10.1103/PhysRevLett.115.126401} {\bibfield
  {journal} {\bibinfo  {journal} {Phys. Rev. Lett.}\ }\textbf {\bibinfo
  {volume} {115}},\ \bibinfo {pages} {126401} (\bibinfo {year}
  {2015})}\BibitemShut {NoStop}%
\bibitem [{\citenamefont {Murthy}\ and\ \citenamefont
  {Shankar}(2012)}]{murthyshankar2012}%
  \BibitemOpen
  \bibfield  {author} {\bibinfo {author} {\bibfnamefont {G.}~\bibnamefont
  {Murthy}}\ and\ \bibinfo {author} {\bibfnamefont {R.}~\bibnamefont
  {Shankar}},\ }\href {\doibase 10.1103/PhysRevB.86.195146} {\bibfield
  {journal} {\bibinfo  {journal} {Phys. Rev. B}\ }\textbf {\bibinfo {volume}
  {86}},\ \bibinfo {pages} {195146} (\bibinfo {year} {2012})}\BibitemShut
  {NoStop}%
\bibitem [{\citenamefont {Sohal}\ \emph {et~al.}(2018)\citenamefont {Sohal},
  \citenamefont {Santos},\ and\ \citenamefont {Fradkin}}]{Sohal-2018}%
  \BibitemOpen
  \bibfield  {author} {\bibinfo {author} {\bibfnamefont {R.}~\bibnamefont
  {Sohal}}, \bibinfo {author} {\bibfnamefont {L.~H.}\ \bibnamefont {Santos}}, \
  and\ \bibinfo {author} {\bibfnamefont {E.}~\bibnamefont {Fradkin}},\ }\href
  {\doibase 10.1103/PhysRevB.97.125131} {\bibfield  {journal} {\bibinfo
  {journal} {Phys. Rev. B}\ }\textbf {\bibinfo {volume} {97}},\ \bibinfo
  {pages} {125131} (\bibinfo {year} {2018})}\BibitemShut {NoStop}%
\bibitem [{\citenamefont {Lu}\ \emph {et~al.}(2020)\citenamefont {Lu},
  \citenamefont {Ran},\ and\ \citenamefont {Oshikawa}}]{Lu-Ran-Oshikawa2020}%
  \BibitemOpen
  \bibfield  {author} {\bibinfo {author} {\bibfnamefont {Y.-M.}\ \bibnamefont
  {Lu}}, \bibinfo {author} {\bibfnamefont {Y.}~\bibnamefont {Ran}}, \ and\
  \bibinfo {author} {\bibfnamefont {M.}~\bibnamefont {Oshikawa}},\ }\href
  {\doibase https://doi.org/10.1016/j.aop.2019.168060} {\bibfield  {journal}
  {\bibinfo  {journal} {Annals of Physics}\ }\textbf {\bibinfo {volume}
  {413}},\ \bibinfo {pages} {168060} (\bibinfo {year} {2020})}\BibitemShut
  {NoStop}%
\bibitem [{\citenamefont {Pfannkuche}\ and\ \citenamefont
  {MacDonald}(1997)}]{pfannkuche1997}%
  \BibitemOpen
  \bibfield  {author} {\bibinfo {author} {\bibfnamefont {D.}~\bibnamefont
  {Pfannkuche}}\ and\ \bibinfo {author} {\bibfnamefont {A.~H.}\ \bibnamefont
  {MacDonald}},\ }\href@noop {} {\bibfield  {journal} {\bibinfo  {journal}
  {Physical Review B}\ }\textbf {\bibinfo {volume} {56}},\ \bibinfo {pages}
  {R7100} (\bibinfo {year} {1997})}\BibitemShut {NoStop}%
\bibitem [{\citenamefont {Sato}\ \emph {et~al.}(2008)\citenamefont {Sato},
  \citenamefont {Tobe},\ and\ \citenamefont {Kohmoto}}]{sato-PRB-2008}%
  \BibitemOpen
  \bibfield  {author} {\bibinfo {author} {\bibfnamefont {M.}~\bibnamefont
  {Sato}}, \bibinfo {author} {\bibfnamefont {D.}~\bibnamefont {Tobe}}, \ and\
  \bibinfo {author} {\bibfnamefont {M.}~\bibnamefont {Kohmoto}},\ }\href
  {\doibase 10.1103/PhysRevB.78.235322} {\bibfield  {journal} {\bibinfo
  {journal} {Phys. Rev. B}\ }\textbf {\bibinfo {volume} {78}},\ \bibinfo
  {pages} {235322} (\bibinfo {year} {2008})}\BibitemShut {NoStop}%
\bibitem [{\citenamefont {Lee}\ \emph {et~al.}(2018)\citenamefont {Lee},
  \citenamefont {Wang}, \citenamefont {Zaletel}, \citenamefont {Vishwanath},\
  and\ \citenamefont {He}}]{Lee-PRX-2018}%
  \BibitemOpen
  \bibfield  {author} {\bibinfo {author} {\bibfnamefont {J.~Y.}\ \bibnamefont
  {Lee}}, \bibinfo {author} {\bibfnamefont {C.}~\bibnamefont {Wang}}, \bibinfo
  {author} {\bibfnamefont {M.~P.}\ \bibnamefont {Zaletel}}, \bibinfo {author}
  {\bibfnamefont {A.}~\bibnamefont {Vishwanath}}, \ and\ \bibinfo {author}
  {\bibfnamefont {Y.-C.}\ \bibnamefont {He}},\ }\href {\doibase
  10.1103/PhysRevX.8.031015} {\bibfield  {journal} {\bibinfo  {journal} {Phys.
  Rev. X}\ }\textbf {\bibinfo {volume} {8}},\ \bibinfo {pages} {031015}
  (\bibinfo {year} {2018})}\BibitemShut {NoStop}%
\bibitem [{\citenamefont {Jain}\ \emph {et~al.}(1990)\citenamefont {Jain},
  \citenamefont {Kivelson},\ and\ \citenamefont
  {Trivedi}}]{Jain-Kivelson-Trivedi-1990}%
  \BibitemOpen
  \bibfield  {author} {\bibinfo {author} {\bibfnamefont {J.~K.}\ \bibnamefont
  {Jain}}, \bibinfo {author} {\bibfnamefont {S.~A.}\ \bibnamefont {Kivelson}},
  \ and\ \bibinfo {author} {\bibfnamefont {N.}~\bibnamefont {Trivedi}},\ }\href
  {\doibase 10.1103/PhysRevLett.64.1297} {\bibfield  {journal} {\bibinfo
  {journal} {Phys. Rev. Lett.}\ }\textbf {\bibinfo {volume} {64}},\ \bibinfo
  {pages} {1297} (\bibinfo {year} {1990})}\BibitemShut {NoStop}%
\bibitem [{\citenamefont {Kivelson}\ \emph {et~al.}(1992)\citenamefont
  {Kivelson}, \citenamefont {Lee},\ and\ \citenamefont
  {Zhang}}]{Kivelson-Lee-Zhang-1992}%
  \BibitemOpen
  \bibfield  {author} {\bibinfo {author} {\bibfnamefont {S.}~\bibnamefont
  {Kivelson}}, \bibinfo {author} {\bibfnamefont {D.-H.}\ \bibnamefont {Lee}}, \
  and\ \bibinfo {author} {\bibfnamefont {S.-C.}\ \bibnamefont {Zhang}},\ }\href
  {\doibase 10.1103/PhysRevB.46.2223} {\bibfield  {journal} {\bibinfo
  {journal} {Phys. Rev. B}\ }\textbf {\bibinfo {volume} {46}},\ \bibinfo
  {pages} {2223} (\bibinfo {year} {1992})}\BibitemShut {NoStop}%
\bibitem [{\citenamefont {Gibertini}\ \emph {et~al.}(2009)\citenamefont
  {Gibertini}, \citenamefont {Singha}, \citenamefont {Pellegrini},
  \citenamefont {Polini}, \citenamefont {Vignale}, \citenamefont {Pinczuk},
  \citenamefont {Pfeiffer},\ and\ \citenamefont {West}}]{Gibertini2009}%
  \BibitemOpen
  \bibfield  {author} {\bibinfo {author} {\bibfnamefont {M.}~\bibnamefont
  {Gibertini}}, \bibinfo {author} {\bibfnamefont {A.}~\bibnamefont {Singha}},
  \bibinfo {author} {\bibfnamefont {V.}~\bibnamefont {Pellegrini}}, \bibinfo
  {author} {\bibfnamefont {M.}~\bibnamefont {Polini}}, \bibinfo {author}
  {\bibfnamefont {G.}~\bibnamefont {Vignale}}, \bibinfo {author} {\bibfnamefont
  {A.}~\bibnamefont {Pinczuk}}, \bibinfo {author} {\bibfnamefont {L.~N.}\
  \bibnamefont {Pfeiffer}}, \ and\ \bibinfo {author} {\bibfnamefont {K.~W.}\
  \bibnamefont {West}},\ }\href {\doibase 10.1103/PhysRevB.79.241406}
  {\bibfield  {journal} {\bibinfo  {journal} {Phys. Rev. B}\ }\textbf {\bibinfo
  {volume} {79}},\ \bibinfo {pages} {241406} (\bibinfo {year}
  {2009})}\BibitemShut {NoStop}%
\bibitem [{\citenamefont {Singha}\ \emph {et~al.}(2011)\citenamefont {Singha},
  \citenamefont {Gibertini}, \citenamefont {Karmakar}, \citenamefont {Yuan},
  \citenamefont {Polini}, \citenamefont {Vignale}, \citenamefont {Katsnelson},
  \citenamefont {Pinczuk}, \citenamefont {Pfeiffer}, \citenamefont {West},\
  and\ \citenamefont {Pellegrini}}]{Singha2011}%
  \BibitemOpen
  \bibfield  {author} {\bibinfo {author} {\bibfnamefont {A.}~\bibnamefont
  {Singha}}, \bibinfo {author} {\bibfnamefont {M.}~\bibnamefont {Gibertini}},
  \bibinfo {author} {\bibfnamefont {B.}~\bibnamefont {Karmakar}}, \bibinfo
  {author} {\bibfnamefont {S.}~\bibnamefont {Yuan}}, \bibinfo {author}
  {\bibfnamefont {M.}~\bibnamefont {Polini}}, \bibinfo {author} {\bibfnamefont
  {G.}~\bibnamefont {Vignale}}, \bibinfo {author} {\bibfnamefont {M.~I.}\
  \bibnamefont {Katsnelson}}, \bibinfo {author} {\bibfnamefont
  {A.}~\bibnamefont {Pinczuk}}, \bibinfo {author} {\bibfnamefont {L.~N.}\
  \bibnamefont {Pfeiffer}}, \bibinfo {author} {\bibfnamefont {K.~W.}\
  \bibnamefont {West}}, \ and\ \bibinfo {author} {\bibfnamefont
  {V.}~\bibnamefont {Pellegrini}},\ }\href {\doibase 10.1126/science.1204333}
  {\bibfield  {journal} {\bibinfo  {journal} {Science}\ }\textbf {\bibinfo
  {volume} {332}},\ \bibinfo {pages} {1176} (\bibinfo {year}
  {2011})}\BibitemShut {NoStop}%
\bibitem [{\citenamefont {Soibel}\ \emph {et~al.}(1996)\citenamefont {Soibel},
  \citenamefont {Meirav}, \citenamefont {Mahalu},\ and\ \citenamefont
  {Shtrikman}}]{Soibel_1996}%
  \BibitemOpen
  \bibfield  {author} {\bibinfo {author} {\bibfnamefont {A.}~\bibnamefont
  {Soibel}}, \bibinfo {author} {\bibfnamefont {U.}~\bibnamefont {Meirav}},
  \bibinfo {author} {\bibfnamefont {D.}~\bibnamefont {Mahalu}}, \ and\ \bibinfo
  {author} {\bibfnamefont {H.}~\bibnamefont {Shtrikman}},\ }\href {\doibase
  10.1088/0268-1242/11/11/019} {\bibfield  {journal} {\bibinfo  {journal}
  {Semiconductor Science and Technology}\ }\textbf {\bibinfo {volume} {11}},\
  \bibinfo {pages} {1756} (\bibinfo {year} {1996})}\BibitemShut {NoStop}%
\bibitem [{\citenamefont {N{\'{a}}dvorn{\'{\i}}k}\ \emph
  {et~al.}(2012)\citenamefont {N{\'{a}}dvorn{\'{\i}}k}, \citenamefont {Orlita},
  \citenamefont {Goncharuk}, \citenamefont {Smr{\v{c}}ka}, \citenamefont
  {Nov{\'{a}}k}, \citenamefont {Jurka}, \citenamefont {Hru{\v{s}}ka},
  \citenamefont {V{\'{y}}born{\'{y}}}, \citenamefont {Wasilewski},
  \citenamefont {Potemski},\ and\ \citenamefont
  {V{\'{y}}born{\'{y}}}}]{N_dvorn_k_2012}%
  \BibitemOpen
  \bibfield  {author} {\bibinfo {author} {\bibfnamefont {L.}~\bibnamefont
  {N{\'{a}}dvorn{\'{\i}}k}}, \bibinfo {author} {\bibfnamefont {M.}~\bibnamefont
  {Orlita}}, \bibinfo {author} {\bibfnamefont {N.~A.}\ \bibnamefont
  {Goncharuk}}, \bibinfo {author} {\bibfnamefont {L.}~\bibnamefont
  {Smr{\v{c}}ka}}, \bibinfo {author} {\bibfnamefont {V.}~\bibnamefont
  {Nov{\'{a}}k}}, \bibinfo {author} {\bibfnamefont {V.}~\bibnamefont {Jurka}},
  \bibinfo {author} {\bibfnamefont {K.}~\bibnamefont {Hru{\v{s}}ka}}, \bibinfo
  {author} {\bibfnamefont {Z.}~\bibnamefont {V{\'{y}}born{\'{y}}}}, \bibinfo
  {author} {\bibfnamefont {Z.~R.}\ \bibnamefont {Wasilewski}}, \bibinfo
  {author} {\bibfnamefont {M.}~\bibnamefont {Potemski}}, \ and\ \bibinfo
  {author} {\bibfnamefont {K.}~\bibnamefont {V{\'{y}}born{\'{y}}}},\ }\href
  {\doibase 10.1088/1367-2630/14/5/053002} {\bibfield  {journal} {\bibinfo
  {journal} {New Journal of Physics}\ }\textbf {\bibinfo {volume} {14}},\
  \bibinfo {pages} {053002} (\bibinfo {year} {2012})}\BibitemShut {NoStop}%
\bibitem [{\citenamefont {Wang}\ \emph {et~al.}(2016)\citenamefont {Wang},
  \citenamefont {Scarabelli}, \citenamefont {Kuznetsova}, \citenamefont {Wind},
  \citenamefont {Pinczuk}, \citenamefont {Pellegrini}, \citenamefont {Manfra},
  \citenamefont {Gardner}, \citenamefont {Pfeiffer},\ and\ \citenamefont
  {West}}]{Wang2016}%
  \BibitemOpen
  \bibfield  {author} {\bibinfo {author} {\bibfnamefont {S.}~\bibnamefont
  {Wang}}, \bibinfo {author} {\bibfnamefont {D.}~\bibnamefont {Scarabelli}},
  \bibinfo {author} {\bibfnamefont {Y.~Y.}\ \bibnamefont {Kuznetsova}},
  \bibinfo {author} {\bibfnamefont {S.~J.}\ \bibnamefont {Wind}}, \bibinfo
  {author} {\bibfnamefont {A.}~\bibnamefont {Pinczuk}}, \bibinfo {author}
  {\bibfnamefont {V.}~\bibnamefont {Pellegrini}}, \bibinfo {author}
  {\bibfnamefont {M.~J.}\ \bibnamefont {Manfra}}, \bibinfo {author}
  {\bibfnamefont {G.~C.}\ \bibnamefont {Gardner}}, \bibinfo {author}
  {\bibfnamefont {L.~N.}\ \bibnamefont {Pfeiffer}}, \ and\ \bibinfo {author}
  {\bibfnamefont {K.~W.}\ \bibnamefont {West}},\ }\href {\doibase
  10.1063/1.4962461} {\bibfield  {journal} {\bibinfo  {journal} {Applied
  Physics Letters}\ }\textbf {\bibinfo {volume} {109}},\ \bibinfo {pages}
  {113101} (\bibinfo {year} {2016})}\BibitemShut {NoStop}%
\bibitem [{\citenamefont {Wang}\ \emph {et~al.}(2018)\citenamefont {Wang},
  \citenamefont {Scarabelli}, \citenamefont {Du}, \citenamefont {Kuznetsova},
  \citenamefont {Pfeiffer}, \citenamefont {West}, \citenamefont {Gardner},
  \citenamefont {Manfra}, \citenamefont {Pellegrini}, \citenamefont {Wind},\
  and\ \citenamefont {Pinczuk}}]{Wang2018}%
  \BibitemOpen
  \bibfield  {author} {\bibinfo {author} {\bibfnamefont {S.}~\bibnamefont
  {Wang}}, \bibinfo {author} {\bibfnamefont {D.}~\bibnamefont {Scarabelli}},
  \bibinfo {author} {\bibfnamefont {L.}~\bibnamefont {Du}}, \bibinfo {author}
  {\bibfnamefont {Y.~Y.}\ \bibnamefont {Kuznetsova}}, \bibinfo {author}
  {\bibfnamefont {L.~N.}\ \bibnamefont {Pfeiffer}}, \bibinfo {author}
  {\bibfnamefont {K.~W.}\ \bibnamefont {West}}, \bibinfo {author}
  {\bibfnamefont {G.~C.}\ \bibnamefont {Gardner}}, \bibinfo {author}
  {\bibfnamefont {M.~J.}\ \bibnamefont {Manfra}}, \bibinfo {author}
  {\bibfnamefont {V.}~\bibnamefont {Pellegrini}}, \bibinfo {author}
  {\bibfnamefont {S.~J.}\ \bibnamefont {Wind}}, \ and\ \bibinfo {author}
  {\bibfnamefont {A.}~\bibnamefont {Pinczuk}},\ }\href {\doibase
  10.1038/s41565-017-0006-x} {\bibfield  {journal} {\bibinfo  {journal} {Nature
  Nanotechnology}\ }\textbf {\bibinfo {volume} {13}},\ \bibinfo {pages} {29}
  (\bibinfo {year} {2018})}\BibitemShut {NoStop}%
\bibitem [{\citenamefont {Van~Hove}(1953)}]{vHS-1953}%
  \BibitemOpen
  \bibfield  {author} {\bibinfo {author} {\bibfnamefont {L.}~\bibnamefont
  {Van~Hove}},\ }\href {\doibase 10.1103/PhysRev.89.1189} {\bibfield  {journal}
  {\bibinfo  {journal} {Phys. Rev.}\ }\textbf {\bibinfo {volume} {89}},\
  \bibinfo {pages} {1189} (\bibinfo {year} {1953})}\BibitemShut {NoStop}%
\bibitem [{sup()}]{supplmat}%
  \BibitemOpen
  \href@noop {} {}\bibinfo {note} {See Supplemental Material for details on
  further examples of the relation between the vHs energy scale and topological
  phase transitions.}\BibitemShut {Stop}%
\bibitem [{\citenamefont {Rammal}(1985)}]{rammal1985}%
  \BibitemOpen
  \bibfield  {author} {\bibinfo {author} {\bibfnamefont {R.}~\bibnamefont
  {Rammal}},\ }\href@noop {} {\bibfield  {journal} {\bibinfo  {journal}
  {Journal de Physique}\ }\textbf {\bibinfo {volume} {46}},\ \bibinfo {pages}
  {1345} (\bibinfo {year} {1985})}\BibitemShut {NoStop}%
\bibitem [{\citenamefont {Andrei~Bernevig}\ \emph {et~al.}(2006)\citenamefont
  {Andrei~Bernevig}, \citenamefont {Hughes}, \citenamefont {Zhang},
  \citenamefont {Chen},\ and\ \citenamefont {Wu}}]{bernevig-IJMPB-2006}%
  \BibitemOpen
  \bibfield  {author} {\bibinfo {author} {\bibfnamefont {B.}~\bibnamefont
  {Andrei~Bernevig}}, \bibinfo {author} {\bibfnamefont {T.~L.}\ \bibnamefont
  {Hughes}}, \bibinfo {author} {\bibfnamefont {S.-c.}\ \bibnamefont {Zhang}},
  \bibinfo {author} {\bibfnamefont {H.-d.}\ \bibnamefont {Chen}}, \ and\
  \bibinfo {author} {\bibfnamefont {C.}~\bibnamefont {Wu}},\ }\href {\doibase
  10.1142/S0217979206035448} {\bibfield  {journal} {\bibinfo  {journal}
  {International Journal of Modern Physics B}\ }\textbf {\bibinfo {volume}
  {20}},\ \bibinfo {pages} {3257} (\bibinfo {year} {2006})}\BibitemShut
  {NoStop}%
\bibitem [{\citenamefont {Agazzi}\ \emph {et~al.}(2014)\citenamefont {Agazzi},
  \citenamefont {Eckmann},\ and\ \citenamefont {Graf}}]{Agazzi2014}%
  \BibitemOpen
  \bibfield  {author} {\bibinfo {author} {\bibfnamefont {A.}~\bibnamefont
  {Agazzi}}, \bibinfo {author} {\bibfnamefont {J.-P.}\ \bibnamefont {Eckmann}},
  \ and\ \bibinfo {author} {\bibfnamefont {G.~M.}\ \bibnamefont {Graf}},\
  }\href {\doibase 10.1007/s10955-014-0992-0} {\bibfield  {journal} {\bibinfo
  {journal} {Journal of Statistical Physics}\ }\textbf {\bibinfo {volume}
  {156}},\ \bibinfo {pages} {417} (\bibinfo {year} {2014})}\BibitemShut
  {NoStop}%
\bibitem [{\citenamefont {Thouless}(1983)}]{Thouless-1983}%
  \BibitemOpen
  \bibfield  {author} {\bibinfo {author} {\bibfnamefont {D.~J.}\ \bibnamefont
  {Thouless}},\ }\href {\doibase 10.1103/PhysRevB.28.4272} {\bibfield
  {journal} {\bibinfo  {journal} {Phys. Rev. B}\ }\textbf {\bibinfo {volume}
  {28}},\ \bibinfo {pages} {4272} (\bibinfo {year} {1983})}\BibitemShut
  {NoStop}%
\bibitem [{\citenamefont {Castro~Neto}\ \emph {et~al.}(2009)\citenamefont
  {Castro~Neto}, \citenamefont {Guinea}, \citenamefont {Peres}, \citenamefont
  {Novoselov},\ and\ \citenamefont {Geim}}]{graphene-RMP}%
  \BibitemOpen
  \bibfield  {author} {\bibinfo {author} {\bibfnamefont {A.~H.}\ \bibnamefont
  {Castro~Neto}}, \bibinfo {author} {\bibfnamefont {F.}~\bibnamefont {Guinea}},
  \bibinfo {author} {\bibfnamefont {N.~M.~R.}\ \bibnamefont {Peres}}, \bibinfo
  {author} {\bibfnamefont {K.~S.}\ \bibnamefont {Novoselov}}, \ and\ \bibinfo
  {author} {\bibfnamefont {A.~K.}\ \bibnamefont {Geim}},\ }\href {\doibase
  10.1103/RevModPhys.81.109} {\bibfield  {journal} {\bibinfo  {journal} {Rev.
  Mod. Phys.}\ }\textbf {\bibinfo {volume} {81}},\ \bibinfo {pages} {109}
  (\bibinfo {year} {2009})}\BibitemShut {NoStop}%
\bibitem [{\citenamefont {Rhim}\ and\ \citenamefont {Park}(2012)}]{Rhim12}%
  \BibitemOpen
  \bibfield  {author} {\bibinfo {author} {\bibfnamefont {J.-W.}\ \bibnamefont
  {Rhim}}\ and\ \bibinfo {author} {\bibfnamefont {K.}~\bibnamefont {Park}},\
  }\href {\doibase 10.1103/PhysRevB.86.235411} {\bibfield  {journal} {\bibinfo
  {journal} {Phys. Rev. B}\ }\textbf {\bibinfo {volume} {86}},\ \bibinfo
  {pages} {235411} (\bibinfo {year} {2012})}\BibitemShut {NoStop}%
\bibitem [{\citenamefont {Karnaukhov}(2019)}]{Karnaukhov2019}%
  \BibitemOpen
  \bibfield  {author} {\bibinfo {author} {\bibfnamefont {I.~N.}\ \bibnamefont
  {Karnaukhov}},\ }\href {\doibase
  https://doi.org/10.1016/j.physleta.2019.04.010} {\bibfield  {journal}
  {\bibinfo  {journal} {Physics Letters A}\ }\textbf {\bibinfo {volume}
  {383}},\ \bibinfo {pages} {2114 } (\bibinfo {year} {2019})}\BibitemShut
  {NoStop}%
\bibitem [{\citenamefont {Das}\ \emph {et~al.}(2020)\citenamefont {Das},
  \citenamefont {Kaul},\ and\ \citenamefont {Murthy}}]{Das2020}%
  \BibitemOpen
  \bibfield  {author} {\bibinfo {author} {\bibfnamefont {A.}~\bibnamefont
  {Das}}, \bibinfo {author} {\bibfnamefont {R.~K.}\ \bibnamefont {Kaul}}, \
  and\ \bibinfo {author} {\bibfnamefont {G.}~\bibnamefont {Murthy}},\ }\href
  {\doibase 10.1103/PhysRevB.101.165416} {\bibfield  {journal} {\bibinfo
  {journal} {Phys. Rev. B}\ }\textbf {\bibinfo {volume} {101}},\ \bibinfo
  {pages} {165416} (\bibinfo {year} {2020})}\BibitemShut {NoStop}%
\bibitem [{\citenamefont {Hasegawa}\ \emph {et~al.}(2006)\citenamefont
  {Hasegawa}, \citenamefont {Konno}, \citenamefont {Nakano},\ and\
  \citenamefont {Kohmoto}}]{hasegawaPRB2006}%
  \BibitemOpen
  \bibfield  {author} {\bibinfo {author} {\bibfnamefont {Y.}~\bibnamefont
  {Hasegawa}}, \bibinfo {author} {\bibfnamefont {R.}~\bibnamefont {Konno}},
  \bibinfo {author} {\bibfnamefont {H.}~\bibnamefont {Nakano}}, \ and\ \bibinfo
  {author} {\bibfnamefont {M.}~\bibnamefont {Kohmoto}},\ }\href {\doibase
  10.1103/PhysRevB.74.033413} {\bibfield  {journal} {\bibinfo  {journal} {Phys.
  Rev. B}\ }\textbf {\bibinfo {volume} {74}},\ \bibinfo {pages} {033413}
  (\bibinfo {year} {2006})}\BibitemShut {NoStop}%
\bibitem [{\citenamefont {Wunsch}\ \emph {et~al.}(2008)\citenamefont {Wunsch},
  \citenamefont {Guinea},\ and\ \citenamefont {Sols}}]{WunschNJP2008}%
  \BibitemOpen
  \bibfield  {author} {\bibinfo {author} {\bibfnamefont {B.}~\bibnamefont
  {Wunsch}}, \bibinfo {author} {\bibfnamefont {F.}~\bibnamefont {Guinea}}, \
  and\ \bibinfo {author} {\bibfnamefont {F.}~\bibnamefont {Sols}},\ }\href
  {\doibase 10.1088/1367-2630/10/10/103027} {\bibfield  {journal} {\bibinfo
  {journal} {New Journal of Physics}\ }\textbf {\bibinfo {volume} {10}},\
  \bibinfo {pages} {103027} (\bibinfo {year} {2008})}\BibitemShut {NoStop}%
\bibitem [{\citenamefont {Redlich}(1984)}]{Redlich1984}%
  \BibitemOpen
  \bibfield  {author} {\bibinfo {author} {\bibfnamefont {A.~N.}\ \bibnamefont
  {Redlich}},\ }\href {\doibase 10.1103/PhysRevLett.52.18} {\bibfield
  {journal} {\bibinfo  {journal} {Phys. Rev. Lett.}\ }\textbf {\bibinfo
  {volume} {52}},\ \bibinfo {pages} {18} (\bibinfo {year} {1984})}\BibitemShut
  {NoStop}%
\bibitem [{\citenamefont {Halperin}\ \emph {et~al.}(1993)\citenamefont
  {Halperin}, \citenamefont {Lee},\ and\ \citenamefont {Read}}]{HLR}%
  \BibitemOpen
  \bibfield  {author} {\bibinfo {author} {\bibfnamefont {B.~I.}\ \bibnamefont
  {Halperin}}, \bibinfo {author} {\bibfnamefont {P.~A.}\ \bibnamefont {Lee}}, \
  and\ \bibinfo {author} {\bibfnamefont {N.}~\bibnamefont {Read}},\ }\href
  {\doibase https://doi.org/10.1103/PhysRevB.47.7312} {\bibfield  {journal}
  {\bibinfo  {journal} {Phys. Rev. B}\ }\textbf {\bibinfo {volume} {47}},\
  \bibinfo {pages} {7312} (\bibinfo {year} {1993})}\BibitemShut {NoStop}%
\bibitem [{\citenamefont {Efremov}\ \emph {et~al.}(2019)\citenamefont
  {Efremov}, \citenamefont {Shtyk}, \citenamefont {Rost}, \citenamefont
  {Chamon}, \citenamefont {Mackenzie},\ and\ \citenamefont
  {Betouras}}]{efremov2019}%
  \BibitemOpen
  \bibfield  {author} {\bibinfo {author} {\bibfnamefont {D.~V.}\ \bibnamefont
  {Efremov}}, \bibinfo {author} {\bibfnamefont {A.}~\bibnamefont {Shtyk}},
  \bibinfo {author} {\bibfnamefont {A.~W.}\ \bibnamefont {Rost}}, \bibinfo
  {author} {\bibfnamefont {C.}~\bibnamefont {Chamon}}, \bibinfo {author}
  {\bibfnamefont {A.~P.}\ \bibnamefont {Mackenzie}}, \ and\ \bibinfo {author}
  {\bibfnamefont {J.~J.}\ \bibnamefont {Betouras}},\ }\href@noop {} {\bibfield
  {journal} {\bibinfo  {journal} {Physical Review Letters}\ }\textbf {\bibinfo
  {volume} {123}},\ \bibinfo {pages} {207202} (\bibinfo {year}
  {2019})}\BibitemShut {NoStop}%
\bibitem [{\citenamefont {Yuan}\ \emph {et~al.}(2019)\citenamefont {Yuan},
  \citenamefont {Isobe},\ and\ \citenamefont {Fu}}]{Yuan2019}%
  \BibitemOpen
  \bibfield  {author} {\bibinfo {author} {\bibfnamefont {N.~F.~Q.}\
  \bibnamefont {Yuan}}, \bibinfo {author} {\bibfnamefont {H.}~\bibnamefont
  {Isobe}}, \ and\ \bibinfo {author} {\bibfnamefont {L.}~\bibnamefont {Fu}},\
  }\href {\doibase 10.1038/s41467-019-13670-9} {\bibfield  {journal} {\bibinfo
  {journal} {Nature Communications}\ }\textbf {\bibinfo {volume} {10}},\
  \bibinfo {pages} {5769} (\bibinfo {year} {2019})}\BibitemShut {NoStop}%
\bibitem [{\citenamefont {Herzog-Arbeitman}\ \emph {et~al.}(2020)\citenamefont
  {Herzog-Arbeitman}, \citenamefont {Song}, \citenamefont {Regnault},\ and\
  \citenamefont {Bernevig}}]{Bernevig2020}%
  \BibitemOpen
  \bibfield  {author} {\bibinfo {author} {\bibfnamefont {J.}~\bibnamefont
  {Herzog-Arbeitman}}, \bibinfo {author} {\bibfnamefont {Z.-D.}\ \bibnamefont
  {Song}}, \bibinfo {author} {\bibfnamefont {N.}~\bibnamefont {Regnault}}, \
  and\ \bibinfo {author} {\bibfnamefont {B.~A.}\ \bibnamefont {Bernevig}},\
  }\href {\doibase 10.1103/PhysRevLett.125.236804} {\bibfield  {journal}
  {\bibinfo  {journal} {Phys. Rev. Lett.}\ }\textbf {\bibinfo {volume} {125}},\
  \bibinfo {pages} {236804} (\bibinfo {year} {2020})}\BibitemShut {NoStop}%
\end{thebibliography}%
%%%%%%%%%%%%%%%%%%%%%%%%%%%%%%%%

%%%%%%%%%% Merge with supplemental materials %%%%%%%%%%
\pagebreak
\widetext
\begin{center}
\textbf{\large Supplemental Material}
\end{center}
%%%%%%%%%% Merge with supplemental materials %%%%%%%%%%
%%%%%%%%%% Prefix a "S" to all equations, figures, tables and reset the counter %%%%%%%%%%
\setcounter{equation}{0}
\setcounter{figure}{0}
\setcounter{table}{0}
\setcounter{page}{1}
\makeatletter
\renewcommand{\theequation}{S\arabic{equation}}
\renewcommand{\thefigure}{S\arabic{figure}}
\renewcommand{\bibnumfmt}[1]{[S#1]}
\renewcommand{\citenumfont}[1]{S#1}

\section{Honeycomb Hofstadter Hamiltonian in the momentum space}
In this section we provide details of the momentum space form of the Hamiltonian, which was omitted in the main text.
In the gauge $\bm{A} = \hat{y}(x+\sqrt{3}y)B$, the real space Hamiltionian
\begin{eqnarray}
\label{eq: app - TB Hamiltonian 1}
H = -\sum_{<\bs{r},\bs{r}'>} 
t_{\bs{r},\bs{r}'}
\,
\,
e^
{\mathrm{i}\,
\frac{2\pi}{\phi_{0}}
\,
\int_{\bs{r}}^{\bs{r'}}\,d\bs{x}\cdot\bs{A}(\bs{x})
}
\,
a^{\dagger}_{\bs{r}}
b_{\bs{r}'}
+
\mathrm{H.c.}
\end{eqnarray}
takes the form
\begin{equation}
\label{eq: NNHamiltonian}
H = -\sum_{m,n} 
a^\dagger_{m,n}
\left(
t_3 + \omega^{-n} t_1 \hat{T}_{\bs{a}_1} + \omega^n t_2 \hat{T}_{\bs{a}_2}
\right)
b_{m,n}
+\mathrm{H.c.}
\end{equation}
where 
$t_{1,2,3}$ are nearest neighbor real hopping elements shown in Fig.(1a) of the main text,
$\omega = e^{2\pi\mathrm{i}\frac{p}{q}}$
and $\hat{T}_{\bs{a}_{1,2}}$ are the translation operators
on the $\bs{a}_{1,2}$ directions, with the lattice constant $a=1$. The magnetic unit cell is formed by extending the original one $q$ times in the $\bs{a}_2$ direction (see Fig.(1a) of the main text), giving rise to an effective tight-binding 
description with $2q$ sites per magnetic unit cell, $(a^{s}_{\bs{r}},b^{s}_{\bs{r}})$, where $s = 0, ..., q-1$, and translation vectors 
$
\{\tilde{\bs{a}}_1, \tilde{\bs{a}}_2 \}
=
\{\bs{a}_1, q\,\bs{a}_2 \}
\,.
$
Then in momentum space, we readily find
\begin{equation}
H = 
-
\sum_{\bs{k} \in \textrm{MBZ}}
\psi^{\dagger}_{\bs{k}}
\begin{pmatrix}
0 & h_{\bs{k}}
\\
h^{\dagger}_{\bs{k}} & 0
\end{pmatrix}
\psi_{\bs{k}}
=
-
\sum_{\bs{k} \in \textrm{MBZ}}
\psi^{\dagger}_{\bs{k}}
{\tau_{1}\otimes h_{\bs{k}}}
\psi_{\bs{k}}
\,,
\end{equation}
with nonzero matrix elements of 
the $q \times q$ matrix 
$h_{\bs{k}}$ given by

\begin{equation}
\begin{split}
&\,
\left(h_{\bs{k}}\right)_{ss} = t_3 + t_1\,e^{\mathrm{i}k_{1}}\,\omega^{-s}
\,,\quad
s = 0, ..., q-1
\,,
\\
&\,
\left(h_{\bs{k}}\right)_{s,s+1}
=
t_{2}\,\omega^{s}
\,,\quad
s = 0, ..., q-2
\,,
\\
&\,
\left(h_{\bs{k}}\right)_{q-1,0}
=
t_{2}\,\omega^{q-1}\,e^{\mathrm{i}k_{2}}
\,,
\end{split}{}
\end{equation}
where
$
\bs{k}
=
k_1\,
\tilde{\bs{g}}_{1}
+
k_2\,
\tilde{\bs{g}}_{2}
$
, with
$ k_i \in [-\pi, \pi)$,
denotes the momenta inside the magnetic Brillouin zone (MBZ) 
with 
$
\tilde{\bs{g}}_1 = \frac{1}{3}\hat{x}-\frac{1}{\sqrt{3}}\hat{y}
$
and
$
\tilde{\bs{g}}_2 = \frac{1}{q}(\frac{1}{3}\hat{x}+\frac{1}{\sqrt{3}}\hat{y})
$
being the reciprocal vectors in Cartesian coordinates satisfying $\tilde{\bs{a}_i}\cdot \tilde{\bs{g}}_j = \delta_{i j}$.

{We now provide an explicit proof that the momentum dependence of the characteristic polynomial appears only in the $a_0 = -\xi^2 = (-1)^{q-1}\det(H)$ coefficient.
By examining the matrix elements given in (S4), we notice that the dependence on $k_2$ appears only in $(H_{\bm{k}})_{q-1,q}=(h_{\bm{k}})_{q-1,0}$ and $(H_{\bm{k}})_{q,q-1}=(h^{\dagger}_{\bm{k}})_{0,q-1}$. As such, these two elements only contribute to the 
coefficient $a_0 = -\xi^2 = (-1)^{q-1}\det(H)$
of the characteristic polynomial (2a):}

{
\begin{eqnarray}
a_0 && =  (-1)^{q-1}\det(H)
\nonumber\\
&& = -\prod_{s=0}^{q-1}|t_3+t_1\omega^{s} \mathrm{e}^{-i k_1}|^2 - |\mathrm{e}^{i k_2}\prod_{s=0}^{q-1}t_2\omega^{s}|^2 - (\mathrm{e}^{i k_2}\prod_{s=0}^{q-1}t_2\omega^{s}(t_3+t_1\omega^{s} \mathrm{e}^{-i k_1})+ \mathrm{h.c.})
\nonumber\\
&& =-t_3^{2q} - t_1^{2q} -( t_3^q t_1^q (-1)^{q-1}\mathrm{e}^{-i q k_1} + \mathrm{h.c.}) - t_2^{2q} - ( t_2^q t_3^q \mathrm{e}^{i k_2} +  t_2^q t_1^q (-1)^{q-1}\mathrm{e}^{i (k_2-q k_1)} + \mathrm{h.c.})
\nonumber\\
&& = -|t_1^q\mathrm{e}^{i ((q-1)\pi-q k_1)} +t_2^q\mathrm{e}^{-i k_2}+t_3^q|^2
\nonumber\\
&& =  -\xi^2
\end{eqnarray}}
{
Now we transform to another gauge $\bm{A}' = \hat{y}(x-\sqrt{3}y)B$, which extends the honeycomb unit cell in the $\bm{a}_1$ direction instead of $\bm{a}_2$. Consequently, we have exchanged the momentum components $k_1$ and $k_2$ so that, now the only $k_1$-dependent coefficient is $a_0$.
Therefore, we have proven that the only $\bm{k}$-dependent coefficient in (2a) is $a_0 = -\xi^2$.     
The momentum independent coefficients $a_n(\{t_i\})$, $n = 1, ..., q$  can be determined recursively via Faddeev–LeVerrier algorithm. We present our results for $p = 1$, $q = 3, 4, 5$ as examples:}
\\

{
For $p/q=1/3$:
\begin{eqnarray}
\nonumber
&& a_1 = 3 (t_1^4 + t_2^4 + t_3^4 + t_2^2 t_3^2 + t_1^2 t_2^2 + t_1^2 t_3^2) 
\nonumber\\
&& a_2 = - 3 (t_1^2 + t_2^2 + t_3^2) 
\nonumber\\
&& a_3 = 1 
\,.
\end{eqnarray}}

{ 
For $p/q=1/4$:
\begin{eqnarray}
\nonumber
&& a_1 =  (t_1^6 + t_2^6 + t_3^6) + t_1^4 (t_2^2 + t_3^2) +  t_2^4 (t_1^2 +t_3^2) + t_3^4(t_1^2  + t_2^2) + 2 t_1^2 t_2^2 t_3^2  
\nonumber\\
&& a_2 = - 6 (t_1^4 + t_2^4 + t_3^4) - 8( t_1^2 t_2^2 + t_1^2 t_3^2 + t_2^2 t_3^2 )
\nonumber\\
&& a_3 = 4 (t_1^2 + t_2^2 + t_3^2)
\nonumber\\
&& a_4 = -1  
\,.
\end{eqnarray}}

{
For $p/q=1/5$:
\begin{eqnarray}
\nonumber
&& a_1 =  5 (t_1^8 + t_3^8 + t_2^8) + 5 (t_1^6 (t_2^2 + t_3^2) + t_2^6 ( t_3^2 + t_1^2) + t_3^6 (t_1^2 + t_2^2 )) + 5(t_1^4 t_2^4 + t_2^4 t_3^4 + t_1^4 t_3^4) + \frac{15 + 5 \sqrt{5}}{2} t_1^2 t_2^2 t_3^2 (t_1^2 + t_2^2 + t_3^2 )
\nonumber\\
&& a_2 = -10 (t_1^6 + t_2^6 +  t_3^6)-15 (t_1^4 (t_2^2 + t_3^2) + t_2^4 ( t_1^2 + t_3^2) + 
    t_3^4 ( t_1^2 + t_2^2)) - \frac{45 + 5 \sqrt{5}}{2} t_1^2 t_2^2 t_3^2
\nonumber\\
&& a_3 = 10 (t_1^4 + t_2^4 +  t_3^4) + 15 (t_1^2 t_2^2 + t_1^2 t_3^2 + t_2^2 t_3^2)
\nonumber\\
&& a_4 = -5 (t_1^2 + t_2^2 + t_3^2)
\nonumber\\
&& a_5 = 1
\,.
\end{eqnarray}}

\section{Properties of the Thouless Function}

\begin{figure}[H]
\centering
\includegraphics[width = .5\textwidth]{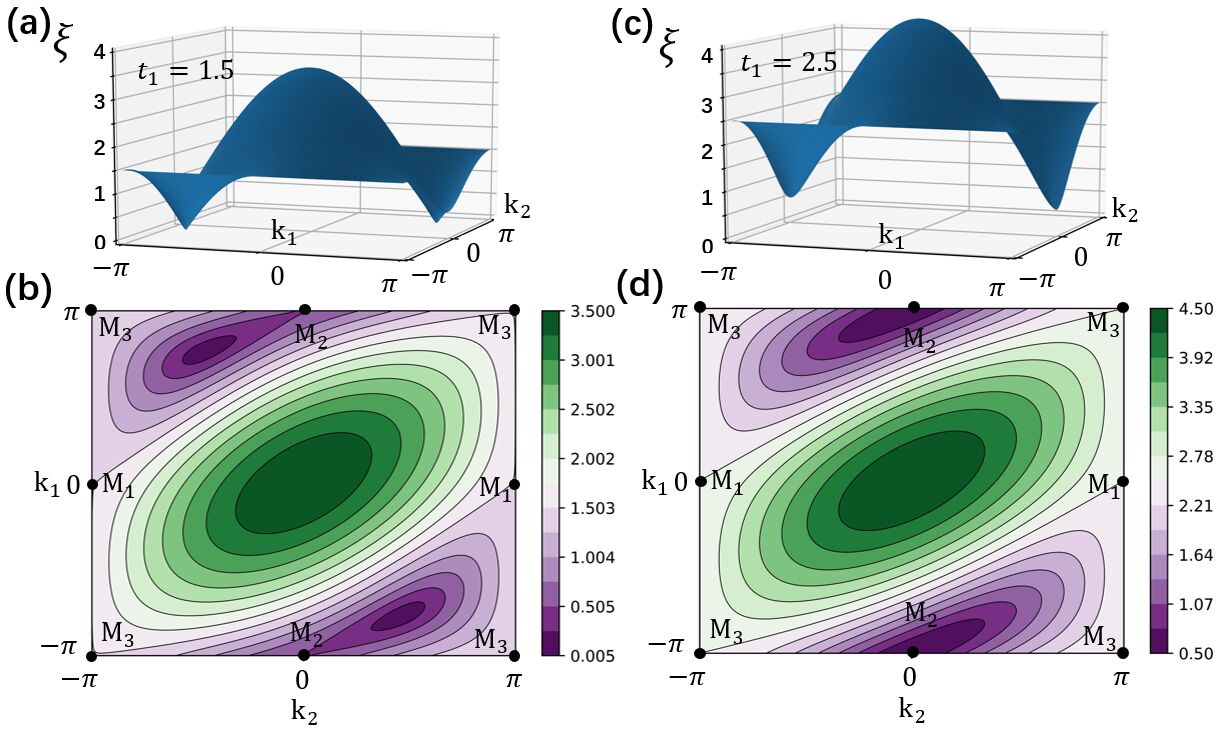}
\caption
{
(a) Thouless's function $\xi(\bm{k})$ with $q=1$ (or equivalently $B=0$) in the first Brillouin zone at $(t_1,t_2,t_3) = (1.5,1,1)$.
(b) Contour plot of $\xi(\bm{k})$ with $q=1$ at $(t_1,t_2,t_3) = (1.5,1,1)$. $\xi_{min}=0$ at the Dirac points $\bm{k}_{0-}=-(\arccos(-0.75),2\arccos(-0.75))=(-2.41,1.44)$ and $\bm{k}_{0+}=(2.41,-1.44)$.
(c)  $\xi(\bm{k})$ with $q=1$ in the first Brillouin zone at $(t_1,t_2,t_3) = (2.5,1,1)$.
(d) Contour plot of $\xi(\bm{k})$ with $q=1$ at $(t_1,t_2,t_3) = (2.5,1,1)$. $\xi_{min}>0$ at $\bm{M}_2=(-\pi,0)$.
The behavior for $q>1$ follows 
upon replacing $(k_1,k_2) \rightarrow (q\,k_1-\pi(q-1), k_2)$
and $\{t_{i}\} \rightarrow \{t^{q}_{i}\}$.
}
\label{fig: xi VHS}
\end{figure}

In this section we provide more detailed analysis of the Thouless function and the constraints imposed upon the Chern bands.
As discussed after in the main text, the mapping from the Thouless function 
$\xi(\{t_i\}, k_1, k_2)$
to the band energies $\{E_{\alpha}(\xi), \alpha = 1, ..., 2q\}$ determines the position of the band extremal and saddle points of all bands for the system with arbitrary rational flux $\phi = \frac{p}{q}\phi_0$. 
We show in Fig.(\ref{fig: xi VHS})  the extremal and saddle points of $\xi(\{t_i\}, k_1, k_2)$ 
for $q=1$ in the two relevant hopping parameter regime $t_2=t_3=1, t_1< 2$ 
and $t_1 > 2$.
The behavior of the energy bands $E_{\alpha}(\xi)$
for $q>1$ can thus be similarly derived  
upon the appropriate replacement $(k_1,k_2) \rightarrow (q\,k_1-\pi(q-1), k_2)$
and $\{t_{i}\} \rightarrow \{t^{q}_{i}\}$.
Moreover, due to particle-hole symmetry, we are led to consider only the bands above charge-neutrality and define the band index $\beta \equiv \alpha-q$.
Therefore, when $0<t_1<2^{1/q}$, the band minimum (maximum) and maximum (minimum) for $\beta$ odd (even) occur, respectively at $\xi(\bm{k}^{(n)}_{0\pm}) = \xi_{min} = 0$ and $\xi(\bm{k}^{(n)}_{max})=\xi_{max} > 0$,
where $\bs{k}^{(n)}_{0\pm}=([\pm\arccos(-t_1^q/2)+\frac{\pi(2n+q-1)}{q}], \pm 2 \arccos(-t_1^q/2))$
%$f_q(t_1)=\arccos(-t_1^q/2)$ 
%
and,
furthermore, $\bm{k}^{(n)}_{max} =
(\pi (2n + q-1)/{q},0)$.
When $t_1>2^{1/q}$, $\xi(\bm{k})>0$ and the minimal of the $\beta$-odd bands and the maximal of the $\beta$-even bands are fixed in $\bm{k}$-space at the $\bm{M}^{(n)}_2$ points.
We can further infer (i) that when $t_1<2^{\frac{1}{q}}$, the isolated energy bands satisfy $dE/d\xi = 0$ at $\xi_{\min} = 0$ so as to prevent the discontinuity in $\nabla E(\bm{k}^{(n)}_{0\pm})$ and (ii) that the $t_1<2^{\frac{1}{q}}$ Chern transitions happen at $\xi(\bm{k}^{(n)}_{max}) = \xi_{max} = |2+t_1^{q}|$, where $|dE/d\xi| \rightarrow \infty$ indicates the presence of $q$ Dirac cones in the points $\bm{k}^{(n)}_{max}$ of the the magnetic Brillouin zone.

{We now address an explanation on why transitions at $(\xi, E) = (0,E_{F} \neq 0)$ are forbidden in case (1) on the bottom of page 2.
Let us begin by considering the band touchings at the particle-hole symmetric points $E=0$. As discussed in the paragraph starting before Eq. (5), $E=0$ band touchings occur
whenever condition Eq. (6) is satisfied.
So, as a consequence of particle-hole symmetry, there is an entire parameter regime where the two center bands touch at $E=0$ forming $2q$ Dirac points localized at the zeros of $\xi$ [the isotropic case was explicitly given in Eq. (5)].
To understand why
$(\xi, E) = (0,E_{F} \neq 0)$ transitions do not occur, 
let us assume otherwise that two bands touch at such a point.
Similarly to the steps discussed in the beginning of the section ``IHCI transitions" (page 2), the characteristic polynomial can be expressed as
$\mathcal{P}(E)
=
c_{2}(E^{2}-E^{2}_{F})^{2} -\xi^{2} + \textrm{O}((E^{2}-E^{2}_{F})^4)
\,,
$
leading to the relation
in the vicinity of
this hypothetical touching point: 
\begin{equation}
\nonumber
\begin{split}
E = \pm \alpha \xi,
\quad
\alpha = (4 c_2 E^{2}_{F})^{-1/2}
\,,
c_2 > 0
\,.
\end{split}    
\end{equation}
So the assumed transition would have the same characteristics of the $2q$ Dirac touchings at $E=0$. However, there is no particle-hole symmetry that protects these $E\neq0$ transitions in parameter regime given by condition (6).
Therefore, we find that the $(\xi, E) = (0,E_{F} \neq 0)$ transitions do not occur when condition (6) holds.}

\section{Further examples of the relation between VHS and Chern transitions}

\begin{figure}[H]
\centering
\includegraphics[width = .8\textwidth]{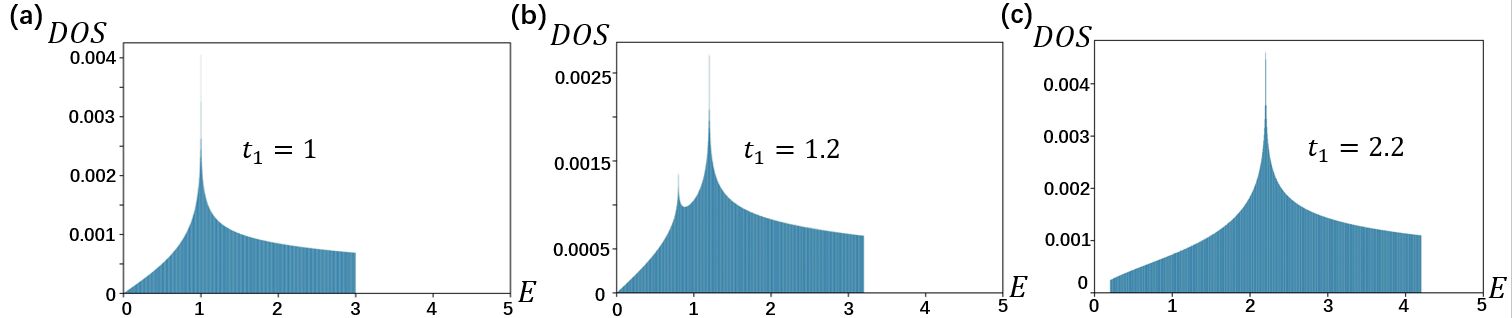}
\caption
{
(a) The $B=0$ density of states at $(t_1,t_2,t_3)=(1,1,1)$.
(b) The $B=0$ density of states at $(t_1,t_2,t_3)=(1.2,1,1)$. The VHS split into two for $0 < t_1 < 2$.
(c) The $B=0$ density of states at $(t_1,t_2,t_3)=(2.2,1,1)$. The Dirac point is gaped and only $E_{\textrm{VHS,1}}$ is left.
}
\label{fig: VHS splitting}
\end{figure}

In this section we present more examples of the relationship between the VHS of the Dirac fermion band near charge neutrality of system A and the onset of topological phase transitions of system B, where the fluxes of A and B are close to each other.
We first recall that the VHS of the $B=0$ (alternatively, $q=1$) graphene band splits into two when the hopping parameter $t_1$ is tuned away from the isotropic lattice $t_1 = t_2 = t_3 = 1$. These VHS energies above charge neutrality,
$E_{\textrm{VHS,1}}$ and $E_{\textrm{VHS,2}}$, 
are given by
\begin{eqnarray}
\label{eq: EVHS}
&& E_{\textrm{VHS,1}} = E(\bm{M}_1) = E(\bm{M}_3) = |t_1|,
\\\nonumber
&& E_{\textrm{VHS,2}} = E(\bm{M}_2) = |2-t_1| \quad (t_1 < 2)
\,.
\end{eqnarray}
If $t_1 > 2$, the Dirac points are gaped and $E_{\textrm{VHS,2}}$ disappears at the lower band edge. As an example, the DOS of the graphene band at different $t_1$ values is shown in Fig.(\ref{fig: VHS splitting}).\par

\begin{figure}[htbp]
\centering
\includegraphics[width = .5\textwidth]{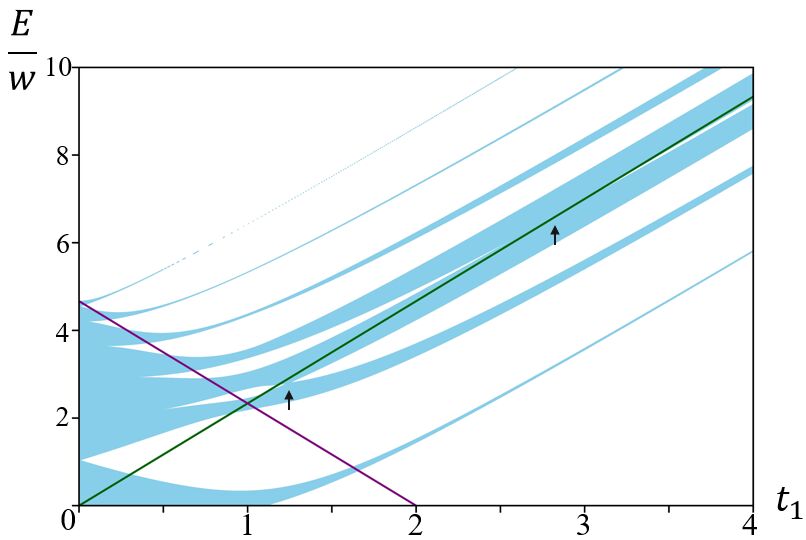}
\caption{
$\phi = 1/7 $ TPTs steered by the $B=0$ VHS's.
The green line shows the van Hove singularity $E_{\textrm{VHS,1}}$ 
and the purple line showing $E_{\textrm{VHS,2}}$. Here $w = 3/7$ is the average energy separation of bands of the $\phi=1/7$ spectrum at $t_1=t_2=t_3=1$. The TPTs are pointed out with black arrows.}
\label{fig: 1-7 phase transitions}
\end{figure}

\begin{figure}[htbp]
\centering
\includegraphics[width = .5\textwidth]{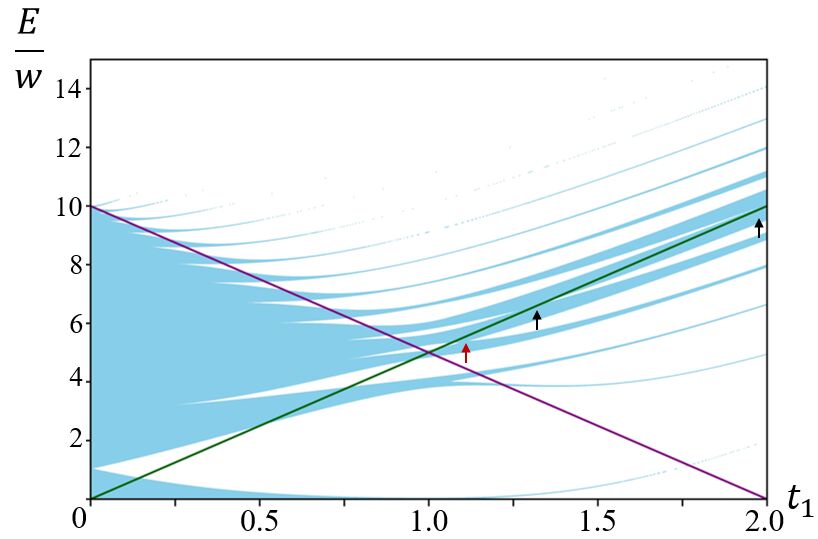}
\caption{
$\phi_{B} = 1/15 $ TPTs steered by the $B=0$ VHS's.
The green line shows the van Hove singularity $E_{\textrm{VHS,1}}$ 
and the purple line showing $E_{\textrm{VHS,2}}$.
Here $w = 1/5$ is the average energy separation of bands of the $\phi=1/15$ spectrum at $t_1=t_2=t_3=1$.
The phase transitions are pointed out with vertical arrows, with the red arrow indicating the composite fermion phase transition.}
\label{fig: 1-15 phase transitions}
\end{figure}

\begin{figure}[htbp]
\centering
\includegraphics[width = .5\textwidth]{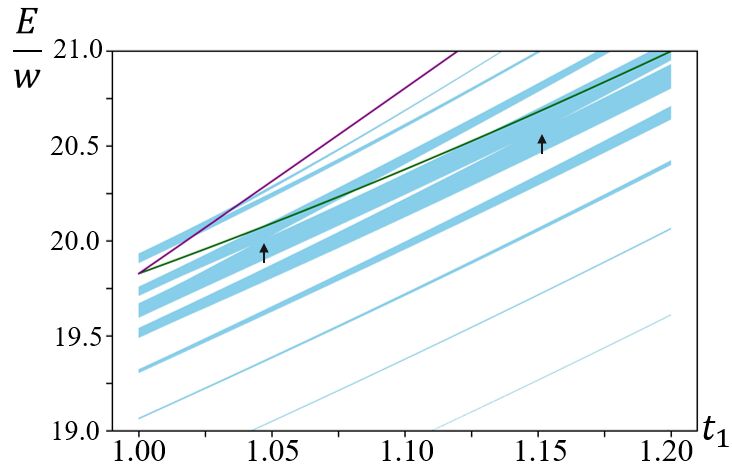}
\caption{
$\phi_B = 12/35 $ TPTs steered by the $\phi_A=1/3$ VHS's of the 2nd band above charge neutrality as an example of the non-Dirac steering.
The green line shows the van Hove singularity $E_{\textrm{VHS,1}}$ 
and the purple line showing $E_{\textrm{VHS,2}}$.
Here $w = 3/35$ is the average energy separation of bands of the $\phi_{B}=12/15$ spectrum at $t_1=t_2=t_3=1$.
The TPT is pointed out with the black arrow.
Notice that the band touchings deviates a litte bit from $E_\textrm{VHS,1}$ suggesting a more complex mechanism behind the non-Dirac TPTs, presumably associated with a narrower band.
}
\label{fig: 12-35 phase transitions}
\end{figure}

The pattern of VHS splitting due to tuning the hopping parameters also persists to the Chern bands as a consequence of the implicit momentum dependence on Thouless function. 
In fact, 
the phenomenon associated with the VHS steering of Chern bands is observed either when $\phi_{A} = 0$
or when there is a nontrivial flux per unit cell.
As argued in the main text, the following conditions must be satisfied: (a) For the steered system B (which undergoes Chern transitions) and the steering system A (which provides the ``background" van Hove singularities), $[(\phi_B-\phi_A)/\phi_0] \ll 1$ and (b) $q_B \gnapprox q_A$. 
We can understand the phase transitions of the system with
$\phi_B=1/q_B$ (with $q_B>1$), 
in terms of the VHS of system A in zero magnetic field. Below, we give concrete examples, one in
Fig.(\ref{fig: 1-7 phase transitions}), which shows the Chern transitions of the $\phi_{B} = 1/7$ system and the second one in Fig.(\ref{fig: 1-15 phase transitions}), which corresponds to a composite fermion transition at $\phi_{CF} = 1/15$ and $n=19/30$. 
In both cases, we see a clear relationship between the onset of phase transitions and the VHS of the $B=0$ band.
Furthermore, we point out that the 
VHS steering mechanism extends to some degree beyond the Dirac bands near charge neutrality discussed in the main text, as shown in Fig.(\ref{fig: 12-35 phase transitions}) for $\phi_{A} = 1/3$ and $\phi_B = 12/35$, albeit the mechanism for higher bands seems more sensitive to the narrow bandwidth of the Chern bands.

\end{document}